\documentclass[review,3p,times]{elsarticle}

\usepackage{amssymb}

\usepackage{hyperref}
\usepackage{graphicx}
\usepackage{dblfloatfix}
\usepackage{float}
\usepackage{graphicx}
\usepackage{times}
\usepackage{epsfig}
\usepackage{bm}
\usepackage{breakurl}
\usepackage{graphicx, subfigure}
\usepackage{amsmath,amssymb, amsfonts, euscript, mathrsfs}
\usepackage{subeqnarray}
\usepackage{cases}
\usepackage{algorithm}
\usepackage{algorithmic}
\usepackage{fancybox}
\usepackage[table]{xcolor}

\definecolor{lightgray}{gray}{0.9}
\definecolor{lightblue}{rgb}{0.98,0.98,1.0}

\usepackage[normalem]{ulem}
\newcommand{\stkout}[1]{\ifmmode\text{\sout{\ensuremath{#1}}}\else\sout{#1}\fi}

\newcommand{\executeiffilenewer}[3]{%
\ifnum\pdfstrcmp{\pdffilemoddate{#1}}%
{\pdffilemoddate{#2}}>0%
{\immediate\write18{#3}}\fi%
}

\newcommand{\T}{\mathsf{T}}

\usepackage{lineno}

\let\oldequation\equation
\let\oldendequation\endequation

\renewenvironment{equation}
  {\linenomathNonumbers\oldequation}
  {\oldendequation\endlinenomath}

\journal{}
\begin{document}

\begin{frontmatter}

\title{Fabrication Sequence Optimization for Minimizing Distortion in Multi-Axis Additive Manufacturing}

\author[Dalian,fn1]{Weiming Wang}
\ead{wwmdlut@dlut.edu.cn}

\author[PME]{Fred van Keulen}
\ead{A.vanKeulen@tudelft.nl}

\author[SDE]{Jun Wu\corref{mycorrespondingauthor}}
\cortext[mycorrespondingauthor]{Corresponding author}
\ead{J.Wu-1@tudelft.nl}

\address[Dalian]{School of Mathematical Sciences, Dalian University of Technology, Dalian, China}
\address[PME]{Department of Precision and
Microsystems Engineering, Delft University of Technology, Delft, The Netherlands}
\address[SDE]{Department of Sustainable Design Engineering, Delft University of Technology, Delft, The Netherlands}

\fntext[fn1]{This work was performed when the first author worked at the Department of Sustainable Design Engineering, Delft University of Technology.}

\begin{abstract}
Additive manufacturing of metal parts involves phase transformations and high temperature gradients which lead to uneven thermal expansion and contraction, and, consequently, distortion of the fabricated components. The distortion has a great influence on the structural performance and dimensional accuracy, e.g., for assembly. It is therefore of critical importance to model, predict and, ultimately, reduce distortion.
In this paper, we present a computational framework for fabrication sequence optimization to minimize distortion in multi-axis additive manufacturing (e.g., robotic wire arc additive manufacturing), in which the fabrication sequence is not limited to planar layers only. We encode the fabrication sequence by a continuous pseudo-time field, and optimize it using gradient-based numerical optimization. To demonstrate this framework, we adopt a computationally tractable yet reasonably accurate model to mimic the material shrinkage in metal additive manufacturing and thus to predict the distortion of the fabricated components. Numerical studies show that optimized curved layers can reduce distortion by orders of magnitude as compared to their planar counterparts.
\end{abstract}


\end{frontmatter}



\section{Introduction}

Multi-axis additive manufacturing using robotic systems can reorient the print head or the component under construction, in addition to provide translational motion~\cite{williams2016wire+,ding2011thermo}. This manufacturing flexibility creates new opportunities and challenges in process planning. In conventional additive manufacturing, which only allows for translation, the fabrication sequence is limited to the use of planar layers. Given a fixed orientation of the component, the stacking of successive planar layers is uniquely determined. In multi-axis additive manufacturing, however, depositing material along curved layers becomes possible~\cite{dai2018support,Fang2020ToG}. The curved layers of a component can be interpreted as an ordered decomposition of the component according to a fabrication sequence. Obviously, the number of possible decompositions is huge. Optimizing the fabrication sequence for multi-axis additive manufacturing offers a new angle to address some of the demanding issues in achieving the highest quality by additive manufacturing. This optimization, however, is a challenging task. It first needs a representation of the fabrication process as optimization variables. To be suitable for gradient-based numerical optimization, this representation is required to be differentiable --- the gradient of a functional with respect to the representation can be calculated. In a recent work~\cite{wang2020space}, we proposed to encode the continuous material deposition process by a pseudo-time field. Each point in the component is associated with a pseudo-time. During the fabrication process, a location with a larger pseudo-time is materialized later than a location with a smaller pseudo-time. The effectiveness of this encoding has been validated on compliance minimization under basic fabrication-process-dependent loads, e.g., gravity~\cite{wang2020space}. The encoding is expected to be applicable to more complex fabrication-process-dependent physics, such as thermomechanical loads and microstructural developments. In the current work, we make an important step towards this end by demonstrating fabrication sequence optimization for minimizing distortion of the fabricated components.

An application scenario that drives our current work is the thermomechanical distortion in robotic wire arc additive manufacturing (WAAM). In WAAM, a metal wire is used as feedstock and an electric arc as the energy source~\cite{williams2016wire+,jafari2021wire}. As a technology for producing metal components, it has a high deposition rate, low material and equipment costs, and a large build volume. It has great potential for fabricating large metallic parts that are commonly seen in aerospace~\cite{ding2011thermo,zalameda2013thermal}, automotive~\cite{mueller2000experiences}, and maritime industries~\cite{ya2017demand}. For instance, Oak Ridge National Laboratory in the US demonstrated the fabrication of two fully functional excavator arms that are over 2 meters tall~\cite{nycz2017challenges,nycz2021effective}. Dutch company MX3D printed and installed a 12-meter-long bridge, using 4.5 tonnes of stainless steel~\footnote{https://mx3d.com/}. This directed energy deposition process involves phase transformations and high temperature gradients as the metal wire is melted and fused to form a component. This results in non-uniform expansion and contraction of the material, and consequently non-negligible residual stresses as well as distortion of the fabricated components~\cite{williams2016wire+,ding2011thermo}. These issues have large impacts on the structural performance and dimensional accuracy, e.g., for assembly. In the related field of welding, the sequence of joining is found to significantly influence the development of residual stresses and distortion~\cite{romero2016welding,beik2019welding}. For instance, depending on the welding sequence, the distortion in pipes can vary by more than seven times~\cite{sattari2008influence}. Specific to WAAM, investigations of different infill patterns and their printing orders also confirm their influences on the residual stresses and distortion~\cite{ding2015wire,mughal2005deformation,mughal2007mechanical,graf2018thermo}. Mitigating residual stresses and minimizing distortion are important topics of current interest for the deployment of WAAM technologies~\cite{williams2016wire+,nycz2021effective}. 

An important module in our fabrication sequence optimization framework is to predict the distortion of the fabricated component. The modelling and simulation of additive manufacturing processes is itself an active research field~\cite{yang2016finite,Korner2020MMTA,Wei2021PMS,Singh2022MTP}. A multitude of process models have been proposed, with different levels of accuracy and computational complexity. WAAM involves time-dependent thermal and mechanical states along with the successive addition of melted material. This could be modelled by a fully coupled transient thermomechanical finite analysis. Such modelling allows gaining insights into the development of residual stresses and distortion due to the fabrication sequence as well as deposition rates and heat sink characteristics~\cite{mughal2005deformation,denlinger2017mitigation,denlinger2016effect,salonitis2016additive}. Similar to multi-pass welding simulation, WAAM simulation involves two major steps: a thermal analysis where the temperature distribution caused by a moving heat source is determined, and a structural analysis which accounts for a series of sequential load steps calculated in a transient thermal analysis~\cite{chiumenti2010finite}. This coupled thermomechanical analysis however is computationally intensive. It is prohibitive for numerical optimization of the fabrication sequence, since numerical optimization requires simulating the fabrication process at each optimization iteration. An appealing alternative is the inherent strain method~\cite{ueda1975new,murakawa1996prediction}, in which the residual stress and distortion are obtained by linear elastic analysis. Due to the computational advantages, inherent strain methods have been increasingly applied for modelling additive manufacturing processes~\cite{ma2016inherent,munro2019process,liang2019modified,chen2019inherent,prabhune2020fast}. Inherent strain methods have been recently integrated into structural topology optimization for minimizing distortion. For instance, Allaire and Jakab{\v{c}}in~\cite{allaire2018taking} proposed a level-set based topology optimization method that considers thermal residual stresses and thermal deformations during the layer-by-layer fabrication process. Pellens et al.~\cite{pellens2020topology} incorporated an inherent strain method in density-based topology optimization to optimize support structures for limiting the vertical displacement of layers during fabrication. Also building upon density-based topology optimization, Misiun et al.~\cite{misiun2021topology} integrated an inherent strain method to optimize the structural layout under distortion constraints. The inherent strain method has also been used in level-set based topology optimization for controlling distortion~\cite{Miki2021CMAME}.

Our main purpose in this paper is to propose the fabrication sequence as optimization variables for minimizing process-order-dependent structural distortion. To this end, we adopt the inherent strain method for its computational efficiency and adapt it to the context of multi-axis additive manufacturing. It is chosen as an example to validate the effectiveness of our fabrication sequence optimization framework. More sophisticated process models are expected to work as well with our framework but are computationally expensive. Our current validation considers some essential manufacturability constraints but is not meant to be comprehensive. Further manufacturability constraints, e.g., regarding the layer thickness, and interference between the print head and already deposited part, constitute future work in the direction opened up by the proposed framework.

The remainder of this paper is organized as follows. In Section~\ref{sec:framework} we present the problem formulation and computational framework. We discuss 2D and 3D numerical examples in Section~\ref{sec:results}, and conclude the paper in Section~\ref{sec:conclusion} with ideas for future work.

\section{Fabrication Sequence Optimization for Distortion Minimization}
\label{sec:framework}

In this section, we start with an overview of the proposed framework, and proceed to the encoding of the fabrication sequence (Section~\ref{subsec:layers}). This is followed by a mechanical simulation model (Section~\ref{subsec:simulation}) and essential manufacturability constraints (Section~\ref{subsec:manufacturability}). The formulation of the optimization problem, together with sensitivity analysis, is given in Section~\ref{subsec:formulation}.

\begin{figure*}[ht!]
\centering
\includegraphics[width=\linewidth]{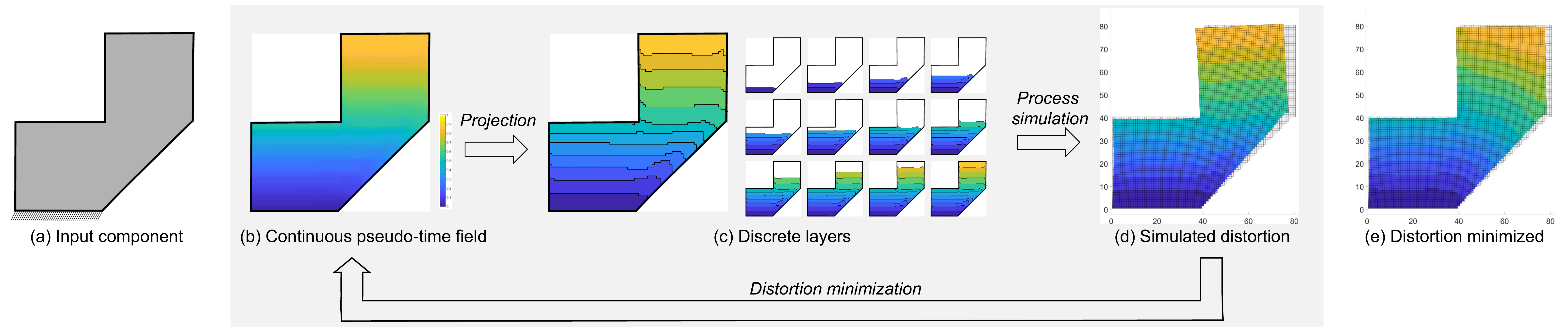}
\caption{The computational framework of fabrication sequence optimization for minimizing distortion. The sequence for an input component~(a) is represented by a continuous pseudo-time field~(b), from which intermediate structures~(c) are calculated by projection operators. The simulated distortion~(d) is reduced by updating the pseudo-time field in an iterative optimization process, leading to minimized distortion~(e).}
\label{fig:workflow}
\end{figure*}

Figure~\ref{fig:workflow} gives an overview of the proposed computational framework. We encode the fabrication sequence using a pseudo-time field (Fig.~\ref{fig:workflow}b). The pseudo-time field is a continuous one, from which a series of intermediate structures is calculated (Fig.~\ref{fig:workflow}c). In this illustration, the component is decomposed by the pseudo-time field into 12 equally-sized layers. The distortion (Fig.~\ref{fig:workflow}d) is predicted by simulating the addition of one layer at a time. In an iterative optimization process, the distortion is minimized by updating the design variables, i.e., the pseudo-time at each point in the domain of the component.

\subsection{Fabrication sequence}
\label{subsec:layers}

We encode the fabrication sequence for additively manufacturing a component by a pseudo-time field, $\bm{t}$. Each point in the domain of the component (specifically, each element in a discretization of the domain) is assigned a pseudo-time variable, $t_e$. 
It shall be emphasized that this pseudo-time field mainly indicates the order of materialization: a larger value indicates an element is to be fabricated later in the process. The pseudo-time field is a normalized one, i.e., $t_e\in[0,1]$.
For simplicity, we also refer to it as the time field. 
From the time field, layers can be extracted by dividing the range of time variables (i.e., $[0,1]$) into uniform intervals, $[T_{j-1}, T_j]$, $j=1,...,N$, with $T_j = \frac{j}{N}$ and $N$ being the number of layers. By evaluating which interval a time value $t_e$ belongs to, it can be decided which layer the corresponding element is part of. The time field thus effectively partitions the component into $N$ layers.

\begin{figure}[tb]
\centering
\def\svgwidth{0.98\linewidth}
\includegraphics[width=0.98\linewidth]{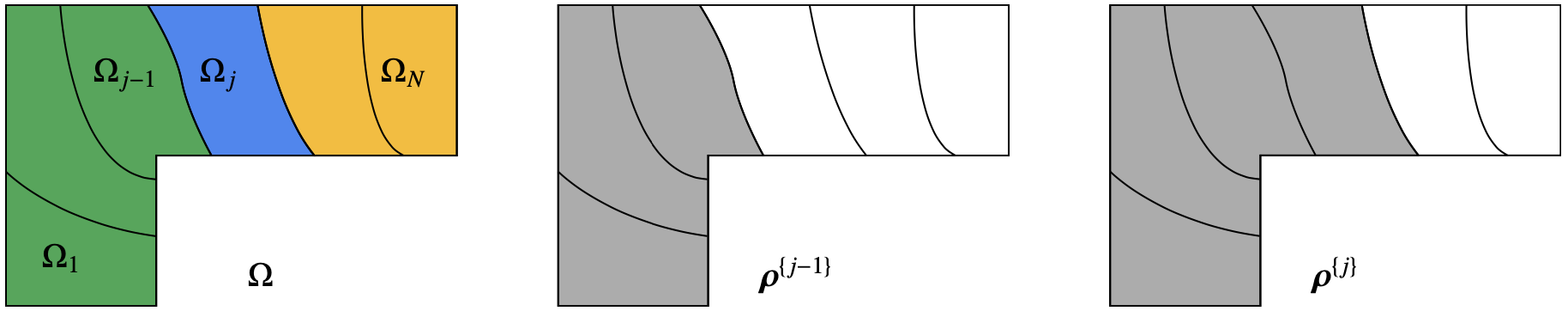}
\caption{Schematic of curved layers for fabricating an $L$-shaped component. Left: Layers of the component. Middle and right: The density field at two consecutive stages.}
\label{fig:layers}
\end{figure}

Consider a component $\Omega$ in 2- or 3-dimensional space. A time field partitions the component into $N$ non-overlapping but successive layers, $\Omega_i$, $i=1,...,N$ (see Fig.~\ref{fig:layers} for illustration on a 2D $L$-shaped component). The layered fabrication process can also be described by a series of pseudo-density fields, $\{\bm{\rho}^{\{0\}},\bm{\rho}^{\{1\}},...,\bm{\rho}^{\{j-1\}},\bm{\rho}^{\{j\}},\bm{\rho}^{\{j+1\}},...,\bm{\rho}^{\{N\}}\}$, where the superscript $\{j\}$ indicates the $j$-th stage during construction, i.e., layers from the first up to the $j$-th have been fabricated. $\bm{\rho}^{\{0\}}$ refers to the beginning of fabrication, $\rho^{\{0\}}_e = 0$, $\forall e$, while $\bm{\rho}^{\{N\}}$ refers to the end of fabrication, $\rho^{\{N\}}_e = 1$,~$\forall e$. At the $j$-th stage, the density of an element depends on the location of its centroid ($x_e$),
\begin{equation}
    \rho^{\{j\}}_e = 
        \begin{cases}
            1, & x_e \in \cup^j_{i=1} \Omega_i, \\
            0, & \text{otherwise.}
        \end{cases}
\label{eq:binaryDensity}
\end{equation}

\begin{figure*}[ht!]
\centering
\includegraphics[width=1\linewidth]{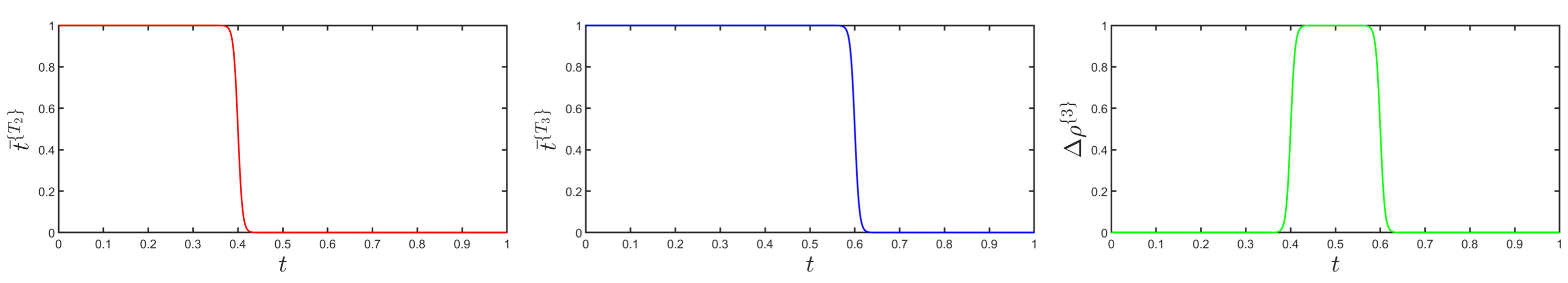}
\caption{Visualization of the projection function, at $T_2=\frac{2}{5}$~(left) and $T_3=\frac{3}{5}$~(middle), and the density of the $3$-rd layer, i.e., $\Delta \bm{\rho}^{\{3\}} =\bar{\bm{t}}^{\{T_3\}} - \bar{\bm{t}}^{\{T_2\}}$~(right).}
\label{fig:projection}
\end{figure*}

To facilitate gradient-based numerical optimization, we present a differentiable operator to partition the component based on the time field. Let $\bm{1}$ denote the density field of a fully solid component, and assume the density field and time field are discretized with the same mesh. The density field consisting of the layers up to the $j$-th one is calculated by an entry-wise multiplication,
\begin{equation}
    \bm{\rho}^{\{j\}} (\bm{t}) = \bm{1} \circ \bar{\bm{t}}^{\{T_j\}}, \quad j=1,...,N.
    \label{eq:rou_t}
\end{equation}
Here $\bar{\bm{t}}^{\{T\}}$ is projected from $\bm{t}$. This projection converts, in a differentiable manner, time values that are smaller than $T$ to $1$, and time values that are larger than $T$ to $0$. This continuous projection is realized by
\begin{equation}
\label{eq:projection}
    \bar{\bm{t}}^{\{T\}} = 1 - \frac{\tanh(\beta_t T)+\tanh(\beta_t(\bm{t} - T))}{\tanh(\beta_t T)+\tanh(\beta_t(1 - T))},
\end{equation}
where $\beta_t$ is a positive number to control the projection sharpness. This sharpness parameter, along with a continuation scheme, is commonly used in density-based topology optimization~\cite{Wang2010SMO,Wu2018TVCG}. Fig.~\ref{fig:projection} (left) and (middle) illustrate the projection function at $T_2=\frac{2}{5}$ and $T_3=\frac{3}{5}$, respectively, where $5$ is the total number of layers. The increment corresponding to the $3$-rd layer is decided by $\Delta \bm{\rho}^{\{3\}} =\bar{\bm{t}}^{\{T_3\}} - \bar{\bm{t}}^{\{T_2\}}$ (right). 

\subsection{Mechanical simulation}
\label{subsec:simulation}

\begin{figure}[!ht]
\centering
\def\svgwidth{0.5\linewidth}
\includegraphics[width=0.5\linewidth]{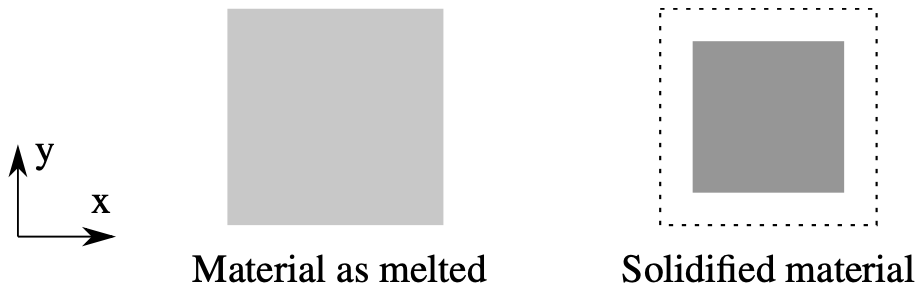}
\caption{Illustration of the shrinkage of material due to an isotropic thermal strain}.
\label{fig:elementContraction}
\end{figure}

Our mechanical simulation is based on an inherent strain method. Imagine a melted material, in isolation from the build plate or previous layers. As it solidifies, its volume shrinks (illustrated in Fig.~\ref{fig:elementContraction}). Using Voigt notation, the thermal strain tensor in 2D is expressed by $\underline{\bm{\varepsilon}}^* = [\epsilon_x,\epsilon_y,\epsilon_{xy}]^{\T}$. The quantities of inherent strains can be determined empirically~\cite{setien2019empirical} or using a detailed process simulation~\cite{Liang2018AM}. The shrinkage of an element due to the thermal strain is equivalently caused by nodal forces ${\bm{f}}$ that correspond to the thermal strain,
\begin{equation}
    {\bm{f}} (\underline{\bm{\varepsilon}}^*) = [\bm{D}]^{\T}[\bm{C}]\underline{\bm{\varepsilon}}^*,
\label{eq:nodalForces}
\end{equation}
where $[\bm{C}]$ denotes the elasticity matrix and $[\bm{D}]$ the strain-displacement relation. 

Extending from an element to the addition of a new layer, the displacement of the simulation grid $\Delta \bm{u}^{\{j\}}$, induced by the thermal strain of elements in the $j$-th layer, $\Omega_j$, can be computed by solving
\begin{equation}
    \bm{K}^{\{j\}} \Delta \bm{u}^{\{j\}} = \bm{f}^{\{j\}}, \quad j = 1,...,N.
\label{eq:equilibriumLayer}
\end{equation}
Here $\bm{K}^{\{j\}}$ is the stiffness matrix for the entire domain $\Omega$, corresponding to the density field $\bm{\rho}^{\{j\}}$. The force vector $\bm{f}^{\{j\}}$ consolidates nodal forces from the thermal strain, $\Delta {\bm{\varepsilon}^*}^{\{j\}}$. The entries of the thermal strain vector $\Delta {\bm{\varepsilon}^*}^{\{j\}}$ are mostly zero, except for elements that are part of $\Omega_j$, i.e.,
\begin{equation}
    \Delta {\bm{\varepsilon}^*_e}^{\{j\}} = 
        \begin{cases}
            \underline{\bm{\varepsilon}}^*, & x_e \in \Omega_j, \\
            [0,0,0]^{\T}, & \text{otherwise.}
        \end{cases}
\label{eq:strain}
\end{equation}

We note that the equilibrium equation Eq.~(\ref{eq:equilibriumLayer}) applies to the entire domain, including both solid elements and the empty grid that has not been filled with material. This is beneficial from two aspects. Firstly, the empty grid is deformed along with the solid elements. We exploit this feature to mimic building a new layer on top of the deformation of previous layers, as in reality in multi-axis additive manufacturing. Secondly, it facilitates implementation as the dimension of equilibrium equations during fabrication is kept constant. The empty elements were referred to as quiet elements in the literature~\cite{michaleris2014modeling}. To avoid singularity of the stiffness matrix due to the inclusion of empty elements, a small yet non-zero density, or equivalently, a minimum stiffness, can be assigned to them. 

With the equilibrium equation (Eq.~(\ref{eq:equilibriumLayer})), we can compute the incremental displacement of the simulation grid due to each addition of a new layer. Under the assumption of small displacements, the displacement field at the $j$-th stage, $\bm{u}^{\{j\}}$, is computed by the summation of incremental displacements until the current stage,
\begin{equation}
    \bm{u}^{\{j\}} = \textstyle{\sum^j_{i=1}}\Delta \bm{u}^{\{i\}}, \quad j = 1,...,N.
\end{equation}
We remark that Eq.~(\ref{eq:equilibriumLayer}) can be solved in parallel for the addition of each layer, since the incremental displacement ($\Delta \bm{u}^{\{j\}}$) induced by the thermal strain of the $j$-th layer ($\Delta {\bm{\varepsilon}^*}^{\{j\}}$) is independent of the thermal strain in prior and latter layers. This follows the analysis of a linearity assumption that was initially proposed for simplifying the simulation of powder-bed processes~\cite{munro2019process} with only planar layers. 

\subsubsection{Mechanical simulation with differentiable layers}

The mechanical simulation model described above assumes a binary description of the layers, i.e., each element belongs exclusively to a single layer. In our framework for fabrication sequence optimization, however, the decomposition into layers is not binary, but is constructed to be differentiable with respect to the time field. In other words, an element may partly belong to multiple layers, with different ratios. The equilibrium equation (Eq.~(\ref{eq:equilibriumLayer})) in mechanical simulation is thus adapted,
\begin{equation}
    \bm{K} (\bm{\rho}^{\{j\}}(\bm{t})) \Delta \bm{u}^{\{j\}} = \bm{f} (\Delta {\bm{\varepsilon}^*}^{\{j\}}(\bm{t})), \quad j = 1,...,N.
    \label{eq:static_eq}
\end{equation}

The stiffness matrix of an element with a density $\rho_e\in[0,1]$ is interpolated from that of a solid element, $\bm{K}_{0}$, by
\begin{equation}
    \label{eq:Young_m}
    \bm{K}_e(\rho_e) = (E_{min} + (1 - E_{min})\rho^p_e)\bm{K}_0,
\end{equation}
where $p$ is a penalization parameter ($p=3$). $E_{min}$ is a small value ($E_{\text{min}} = 10^{-9}$), introduced to prevent the stiffness matrix from becoming singular. 

For the equivalent forces of the thermal strains, on the right-hand side of the equilibrium equation, the conditional function in Eq.~(\ref{eq:strain}) is replaced by
\begin{equation}
    \Delta {\bm{\varepsilon}^*}^{\{j\}}(\bm{t}) |_e= \left(\Delta \bm{\rho}^{\{j\}} (\bm{t})|_e\right)^q  \, \underline{\bm{\varepsilon}}^*.
    \label{eq:loads_proj}
\end{equation}
Here $\underline{\bm{\varepsilon}}^*$ is the thermal strain for an element, as introduced previously. The increment in the density field corresponding to the $j$-th layer is given by
\begin{equation}
    \Delta \bm{\rho}^{\{j\}} (\bm{t}) = \bm{\rho}^{\{j\}} (\bm{t})- \bm{\rho}^{\{j-1\}} (\bm{t}), \quad j=1,...,N,
    \label{eq:inter_t}
\end{equation}
where, as defined previously, $\bm{\rho}^{\{0\}} = \bm{0}$, $\bm{\rho}^{\{N\}} = \bm{1}$. 

In Eq.~(\ref{eq:loads_proj}) the density increment is raised to the power of $q$, similar to the penalization $p$ in the interpolation of stiffness (Eq.~(\ref{eq:Young_m})). If the thermal-strain penalization ($q$) is smaller than the stiffness penalization ($p$), for instance, $p=3$ while $q=1$, the thermal strain creates a large mechanical load for a disproportionally weak material, leading to an excessive displacement. Based on numerical experiments, we suggest having $q\geq p$. In the examples in this paper, we used $q = p$, if not explicitly stated otherwise.

\subsection{Manufacturability constraints}
\label{subsec:manufacturability}

Manufacturability constraints shall be integrated into the optimization to avoid sequences that are infeasible from a manufacturing perspective. Here we present the incorporation of two constraints that we consider essential. Yet we note that these constraints are by no means exhaustive. These constraints were formulated in our prior work of space-time topology optimization~\cite{wang2020space}. In this current work, these constraints are integrated, with some adaptations. We briefly reiterate the key formulation with an emphasis on adaptation. 

\subsubsection{Continuity constraints}

A local minimum in the time field indicates an isolated material patch during the fabrication process; all its adjacent elements have a larger time value and thus have not yet been fabricated. The presence of local minima would necessitate auxiliary structures during fabrication. A local maximum in the time field, if not located on the contour of the component, indicates an enclosed void, which is inaccessible; all its adjacent elements have a smaller time value and thus have already been fabricated.

Isolated material patches and enclosed voids can be prevented by requiring
\begin{equation}
    \min_{i \in \mathcal{N}_e} (t_i) \le t_e \le \max_{i \in \mathcal{N}_e} (t_i), \; \forall e\in \mathcal{M},
\label{eq:minmax}
\end{equation}
where $\mathcal{N}_e$ denotes the set of elements adjacent to element $e$. The set $\mathcal{M}$ includes all elements in the domain except those that are prescribed as the starting point/region for the fabrication process (i.e., with $t_e=0$, typically on the building plate).

The $\min$ and $\max$ functions are not differentiable. A possible approach is to approximate them by the $p$-norm~\cite{wang2020space}. In this work, we use an alternative approach that simplifies implementation. Specifically, we consolidate these two constraints by an approximation,
\begin{equation}
    t_e = \textrm{mean}_{i\in \mathcal{N}_e} (t_i), \; \forall e\in \mathcal{M},
\end{equation}
with 
\begin{equation}
\textrm{mean}_{i\in \mathcal{N}_e} (t_i) = \frac{\sum_{i \in \mathcal{N}_e} t_i}{{n} (\mathcal{N}_e)},
\end{equation} 
here, ${n}$ denotes the number of elements in a set. The mean of numbers in a set is bounded by its minimum and maximum number. Therefore, this approximation is conservative, in the sense that it is a sufficient but not necessary condition of Eq.~(\ref{eq:minmax}). The mean function ensures the smoothness of the time field, and also effectively avoids checkerboard patterns that are typically suppressed by convolution filters in density-based topology optimization.

The $\textrm{mean}$ equality constraint applies to all elements in the domain except the starting element. This leads to a large number of constraints. We consolidate them by
\begin{equation}
g_0(\bm{t})=\frac{1}{{n}(\mathcal{M})}\sum_{e\in \mathcal{M}}||t_e - \textrm{mean}_{i\in \mathcal{N}_e} (t_i)||^2 \leq \gamma_c,
\end{equation}
where $\gamma_c$ is a small constant which is set to $10^{-3}$ in our examples. As $\gamma_c$ approaches 0, this constraint effectively restricts $t_e$ towards the mean value of its neighbors. This constraint function is quadratic, for which sensitivity analysis is straightforward.

\subsubsection{Volume constraints on layers}

For simplicity, we assume a constant fabrication speed, i.e., each layer has the same material volume. Denote the total volume of the component by $V^*$. Since the number of layers is prescribed as $N$, the volume up until the $j$-th layer shall be $\frac{j}{N}V^*$,
\begin{equation}
    V^{\{j\}} (\bm{t}) = {\sum_e} \, \rho_e^{\{j\}}(t_e) \cdot v_e = \frac{j}{N}V^*, \quad j = 1,...,N,
    \label{eq:volume_layer}
\end{equation}
where $v_e$ is the volume of an element. A uniform finite element discretization is used in our implementation, and thus $v_e$ is a constant for all elements.

The above equality constraint is replaced by inequality constraints,
\begin{equation}
    -\gamma_v \le g_j(\bm{t}) = \frac{V^{\{j\}}}{V^*} - \frac{j}{N}  \le 0, \quad j = 1,...,N,
    \label{eq:volume_cons}
\end{equation}
where $\gamma_v$ is a small constant for numerical stability ($10^{-3}$ in our examples). The lower bound is needed in minimizing thermal strain-induced distortion, since otherwise, a distortion of $0$ can be achieved by a trivial solution, $\bm{t}=\bm{0}$.

\subsection{Optimization problem formulation}
\label{subsec:formulation}

The optimization aims to reduce the distortion of the fabricated component. Depending on application scenarios, different distortion measurements can be applied, for instance, the absolute displacement of a node (or a few nodes), or the relative displacement between a few selected nodes, e.g., to maintain a flat edge/surface or to maintain a right angle between edges/surfaces. These various distortions are generically encoded by $\bm{u}^{\T} \bm{Q} \bm{u}$, where $\bm{u} = [\bm{u}^{{\{1\}}^{\T}}, ..., \bm{u}^{{\{N\}}^{\T}}]^{\T}$ encompasses the displacement vector during fabrication, and $\bm{Q}$ is a sparse, symmetric matrix to arrive at different measurements. In the examples shown in this paper, the objective concerns the displacement vector $\bm{u}^{{\{N\}}}$, i.e., measured when the entire component has been fabricated.

Our optimization problem is formulated as
\begin{eqnarray}
\label{eq:formulation_dis}
 & \underset{{\bm{t}}}{\min}  & d (\bm{t}) =  \bm{u}^{\T} \bm{Q} \bm{u} \label{eq:objective}\\
    & s.t. & \bm{K}^{\{i\}} (\bm{t}) \Delta \bm{u}^{\{i\}} = \bm{f}^{\{i\}}(\bm{t}), \; i=1,...,N, \\
    &      & -\gamma_v \le g_i(\bm{t}) \le 0,  \; i=1,2,...,N , 
     \label{eq:volume_const}\\
    &      & g_0(\bm{t}) \leq \gamma_c, \label{eq:continuity} \\
    &      & 0 \le t_e \le 1, \; \forall e.
\end{eqnarray}

This optimization problem is solved by gradient-based numerical optimization. In particular, we use the method of moving asymptotes (MMA)~\cite{svanberg1987method}.

\subsubsection{Sensitivity analysis}

Introducing Lagrange multipliers for the equilibrium equations, $\bm{\lambda}_i$, $i=1,2,...,N$, the Lagrange function for the objection function $d(\bm{t})$ is written as:
\begin{equation}
\begin{aligned}
L(\bm{t})= \bm{u}^{\T} \bm{Q} \bm{u}+
\sum_{i=1}^{i=N} \bm{\lambda}_i^{\T} \left(\bm{K}^{\{i\}} (\bm{t}) \Delta \bm{u}^{\{i\}} - \bm{f}^{\{i\}}(\bm{t})\right).
\end{aligned}
\end{equation}

Then, the sensitivity of the objective function regarding the design variable $t_e$ is given by
\begin{equation}
\begin{aligned}
\frac{\partial L(\bm{t})}{\partial t_e} =& 2\bm{u}^{\T} \bm{Q} \frac{\partial \bm{u}}{\partial t_e}+\sum_{i=1}^{i=N} \bm{\lambda}_i^{\T} \left( \frac{\partial \bm{K}^{\{i\}}(\bm{t})}{\partial t_e} \Delta \bm{u}^{\{i\}}(\bm{t}) +\bm{K}^{\{i\}}(\bm{t}) \frac{\partial \Delta \bm{u}^{\{i\}}(\bm{t})}{\partial t_e} - \frac{\partial \bm{f}^{\{i\}}(\bm{t})}{\partial t_e}\right),\\
= & 2\bm{u}^{\T} \bm{Q} \sum_{i=1}^{i=N} \frac{\partial  \Delta \bm{u}^{\{i\}}(\bm{t})}{\partial t_e} + 
\sum_{i=1}^{i=N} \bm{\lambda}_i^{\T} \left( \frac{\partial \bm{K}^{\{i\}}(\bm{t})}{\partial t_e} \Delta \bm{u}^{\{i\}}(\bm{t}) +\bm{K}^{\{i\}}(\bm{t}) \frac{\partial \Delta \bm{u}^{\{i\}}(\bm{t})}{\partial t_e} - \frac{\partial \bm{f}^{\{i\}}(\bm{t})}{\partial t_e}\right).
\end{aligned}
\end{equation}
We then reorder the expression on the right-hand side,
\begin{equation}
\begin{aligned}
\label{eq:sensitivity_2}
\frac{\partial L(\bm{t})}{\partial t_e} =& \sum_{i=1}^{i=N} \left( 2\bm{u}^{\T} \bm{Q} + \bm{\lambda}_i^{\T}\bm{K}^{\{i\}}(\bm{t}) \right) \frac{\partial \Delta \bm{u}^{\{i\}}(\bm{t})}{\partial t_e} +
\sum_{i=1}^{i=N} \bm{\lambda}_i^{\T} \left( \frac{\partial \bm{K}^{\{i\}}(\bm{t})}{\partial t_e} \Delta \bm{u}^{\{i\}}(\bm{t}) - \frac{\partial \bm{f}^{\{i\}}(\bm{t})}{\partial t_e}\right).
\end{aligned}
\end{equation}
We choose $\bm{\lambda}_i$, satisfying
\begin{equation}
    \bm{K}^{\{i\}}(\bm{t})\bm{\lambda}_i + 2 \bm{Q}^{\T}\bm{u}= 0, \; i=1,2,...,N.
    \label{eq:adjoint}
\end{equation}
This simplifies Eq.~(\ref{eq:sensitivity_2}) to
\begin{equation}
\begin{aligned}
\label{eq:sensitivity_3}
\frac{\partial L(\bm{t})}{\partial t_e} =&  \sum_{i=1}^{i=N} \bm{\lambda}_i^{\T} \left( \frac{\partial \bm{K}^{\{i\}}(\bm{t})}{\partial t_e} \Delta \bm{u}^{\{i\}}(\bm{t}) - \frac{\partial \bm{f}^{\{i\}}(\bm{t})}{\partial t_e}\right).
\end{aligned}
\end{equation}

The stiffness matrix of element $e$ depends solely on its time value. Using the chain rule, the derivative becomes
\begin{equation}
    \frac{\partial\bm{K}^{\{i\}}_e}{\partial t_e} =  \frac{\partial \bm{K}_e(\rho^{\{i\}}_e)}{\partial \rho^{\{i\}}_e} \frac{\partial \rho^{\{i\}}_e}{\partial \bar{t}^{\{i\}}_e} \frac{\partial \bar{t}^{\{i\}}_e}{\partial t_e},
\end{equation}
where $\frac{\partial \bm{K}_e(\rho^{\{i\}}_e)}{\partial \rho^{\{i\}}_e}$, $\frac{\partial \rho^{\{i\}}_e}{\partial \bar{t}^{\{i\}}_e}$, and $\frac{\partial \bar{t}^{\{i\}}_e}{\partial t_e}$ can be derived from Eq.~(\ref{eq:Young_m}), Eq.~(\ref{eq:rou_t}), and Eq.~(\ref{eq:projection}), respectively. Then, $\frac{\partial \bm{K}^{\{i\}}(\bm{t})}{\partial t_e}$ in Eq.~(\ref{eq:sensitivity_3}) is assembled from $\frac{\partial\bm{K}^{\{i\}}_e}{\partial t_e}$ of each element.

The isotropic thermal strain of element $e$ contributes equivalent loads to its four nodes in 2D or eight nodes in 3D. From Eqs.~(\ref{eq:nodalForces}) and~(\ref{eq:loads_proj}), we get
\begin{equation}
      \frac{\partial  \bm{f}^{\{i\}}_e}{\partial t_e} = q[\bm{D}]^{\T}[\bm{C}]\underline{\bm{\varepsilon}}^* \left(\Delta \rho^{\{i\}}_e \right)^{q-1}\frac{\partial \Delta \rho^{\{i\}}_e}{\partial t_e},
      \label{eq:forceGradient}
\end{equation}
where $\frac{\partial \Delta \rho^{\{i\}}_e}{\partial t_e}$ can be derived from Eqs.~(\ref{eq:rou_t}) and~(\ref{eq:inter_t}). Then, $\frac{\partial \bm{f}^{\{i\}}(\bm{t})}{\partial t_e}$ in Eq.~(\ref{eq:sensitivity_3}) is assembled from $\frac{\partial \bm{f}^{\{i\}}_e}{\partial t_e}$ of each element.

The continuity constraints are quadratic functions of $\bm{t}$, while the volume constraints are linear functions. The sensitivities can be derived accordingly, and are omitted here.

\section{Numerical Results}
\label{sec:results}

The proposed methods have been implemented using Matlab. In this section, we report and discuss the numerical results of our tests on 2D and 3D components. We use bilinear quadrilateral elements for 2D, and trilinear hexahedral elements for 3D. For demonstration purposes, the Young's modulus of the material is 1, and the Poisson's ratio is 0.3. We use an isotropic thermal strain in most examples, except in Section~\ref{subsec:anisotropy} where we demonstrate an extension of our framework to take in an anisotropy thermal strain.
The isotropic thermal strain in 2D is set as $\epsilon_x=\epsilon_y=-0.01$ and $\epsilon_{xy}=0.0$. Similarly, in 3D, $\epsilon_i=-0.01, i\in\{x,y,z\}$ and the shear component is 0.

\subsection{2D $L$-shaped component}

\begin{figure*}[htb!]
\centering
\includegraphics[width=1\linewidth]{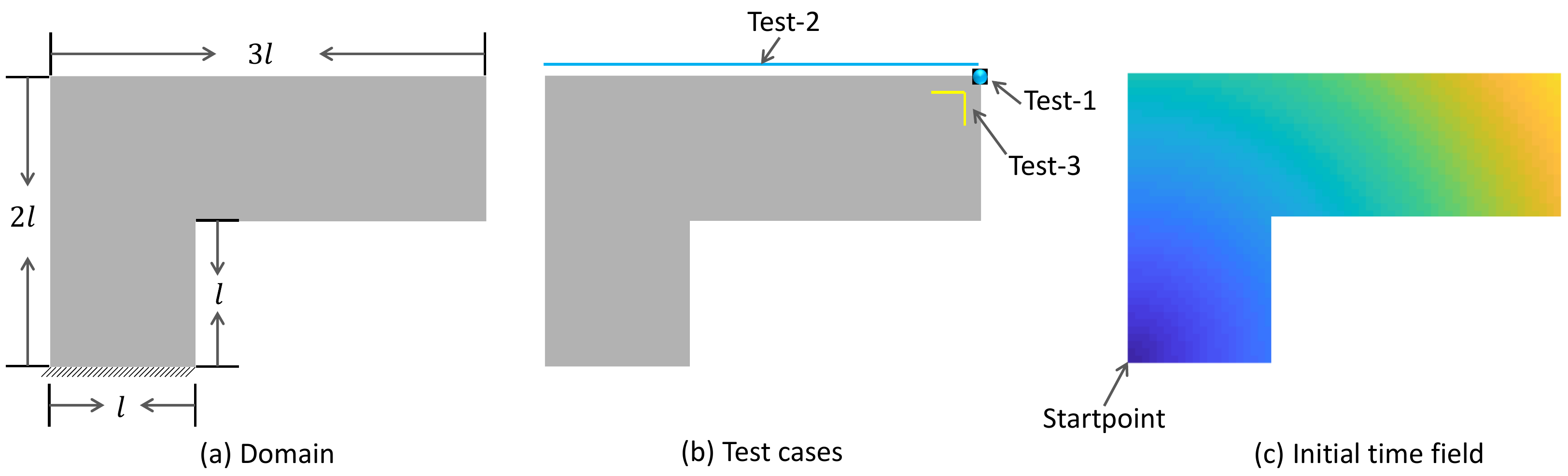}
\caption{(a) Design domain of an $L$-shaped component. (b) The objective is to reduce the distortion, measured by the displacement of a node (Test-1), the flatness of a horizontal edge (Test-2), and the right angle formed by two edges (Test-3). (c) The fabrication starts from the bottom-left element. The time field is initialized by the normalized distance field corresponding to this startpoint.}
\label{fig:L_shape}
\end{figure*}

We first test our framework on a 2D $L$-shaped component (Fig.~\ref{fig:L_shape}a). The domain is discretized by a finite element grid of $120 \times 80$. The nodes on the bottom are fixed, mimicking depositing on a horizontal build plate. We assume the fabrication starts from the element on the bottom left. A distance field corresponding to this element is computed to initialize the design field ($\bm{t}$) (Fig.~\ref{fig:L_shape}c). The prescribed number of layers is $N=8$.

\subsubsection{Distortion measures}

We demonstrate distortion minimization with different distortion measures, as illustrated in Fig.\ref{fig:L_shape}b. This includes minimizing the displacement of a single node (Test-1), maintaining the flatness of an edge (Test-2), and maintaining the perpendicularity of two edges (Test-3). 
\paragraph{Test-1}
The displacement of a node is computed by  
\begin{equation}
    d_1=(u^x_i)^2+(u^y_i)^2,
\end{equation}
where subscript $i$ refers to the index of the node of interest (the top-right node in this test), and the superscripts $x$ and $y$ indicate the $x$ and $y$ components of the displacement vector, respectively. 

\paragraph{Test-2} The flatness of a horizontal edge is computed by
\begin{equation}
    d_2=\frac{1}{{n}(\mathcal{I})}\sum_{i \in \mathcal{I}}\left(u^y_i - \textstyle{\frac{1}{{n}(\mathcal{I})}} \textstyle{\sum_{i \in \mathcal{I}}} u^y_i \right)^2,
    \label{eq:obj-Test2}
\end{equation}
where $\mathcal{I}$ denotes the set of nodes that are sampled from the edge of interest. The number of nodes ${n}(\mathcal{I})$ is equal to or larger than $2$. The flatness is tested with different numbers of sample nodes.

\paragraph{Test-3} The perpendicularity of a horizontal and a vertical edge is measured by the flatness of these two edges,
\begin{equation}
    d_3 = \frac{1}{{n}(\mathcal{I})}\sum_{i \in \mathcal{I}}\left(u^y_i - \textstyle{\frac{1}{{n}(\mathcal{I})}} \textstyle{\sum_{i \in \mathcal{I}}} u^y_i \right)^2 +
    \frac{1}{{n}(\mathcal{J})}\sum_{j \in \mathcal{J}}\left(u^x_j - \textstyle{\frac{1}{{n}(\mathcal{J})}} \textstyle{\sum_{j \in \mathcal{J}}} u^x_j \right)^2,
\end{equation}
where $\mathcal{I}$ and $\mathcal{J}$ denote the set of nodes from the horizontal and vertical edge, respectively. 

\subsubsection{Results}

\begin{figure*}[ht!]
\centering
\includegraphics[width=\linewidth]{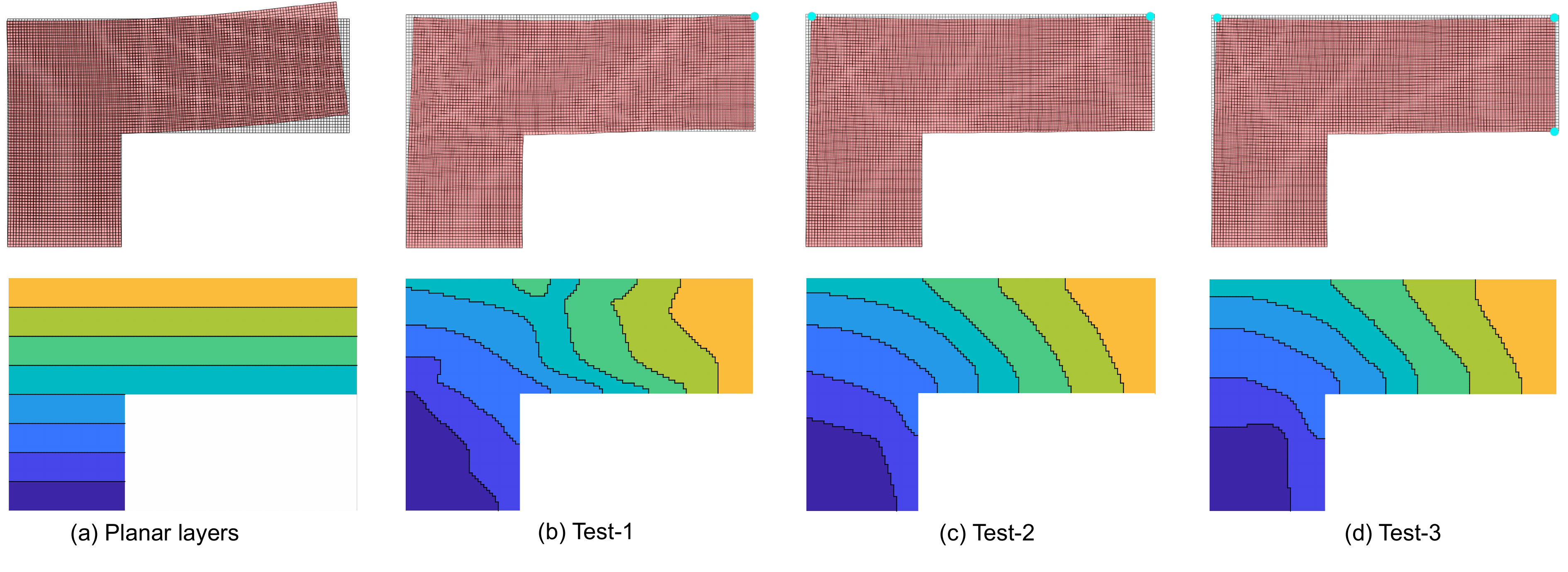}
\caption{(a) Planar layers (bottom) and the corresponding simulated distortion (top). (b-d) The fabrication sequence is optimized to reduce the distortion, with three different distortion measures, i.e., the displacement of the top-right node (Test-1), the flatness of the top edge (Test-2), and the right angle between the top and right edges (Test-3).} \label{fig:Tests}
\end{figure*}

Figure~\ref{fig:Tests} (top) shows the simulated distortion of the fabricated component using different layers (bottom). The background Cartesian grid is superimposed for interpreting the distortion. Comparing with the distortion resulting from horizontal layers (Fig.~\ref{fig:Tests}a), it can be observed that the optimized curved layers lead to a much smaller distortion (b-d). Checking the distortion of interest, one can find out that in Fig.~\ref{fig:Tests}b the top-right node locates close to its reference position in the Cartesian grid, in (c) the two nodes on the top form a straight horizontal line, and in (d) the top edge and right edge, each represented by two nodes, maintain perpendicular to each other. These results confirm the effectiveness of the proposed method.

\begin{figure*}[ht!]
\centering
\includegraphics[width=\linewidth]{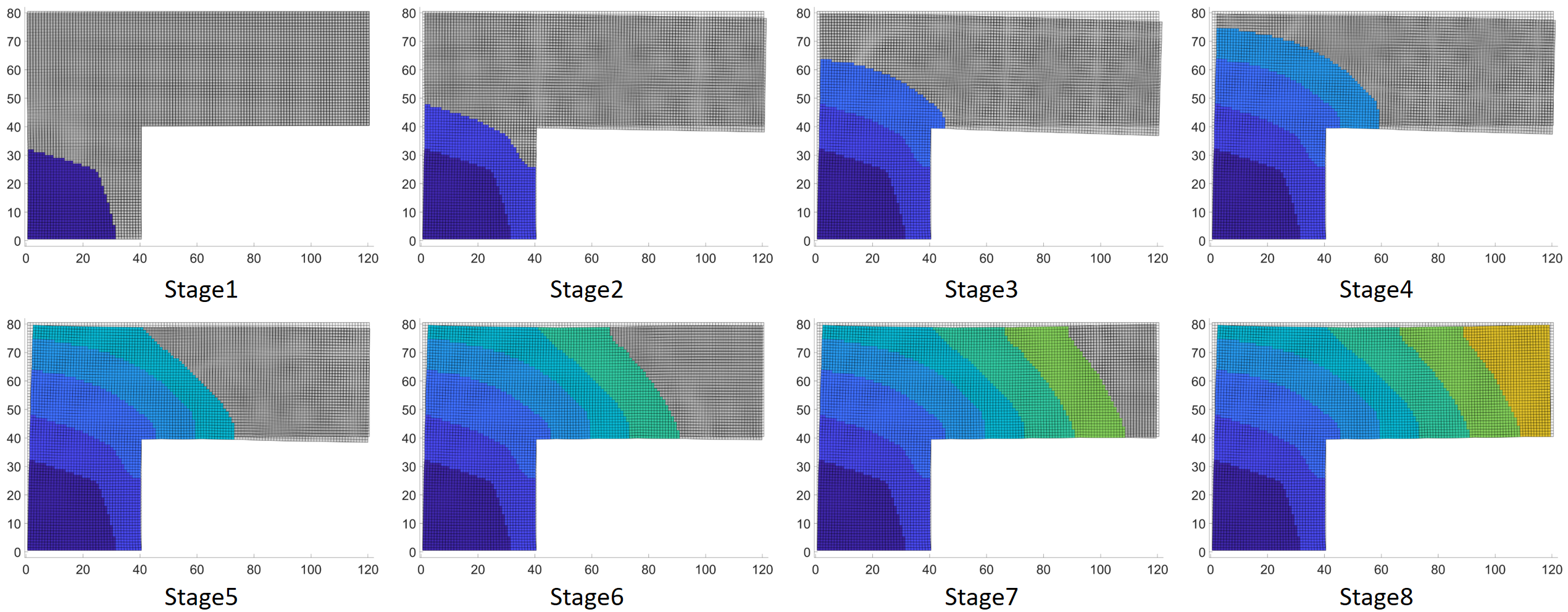}
\caption{Process simulation with the optimized fabrication sequence (Test-2 in Fig.~\ref{fig:Tests}).}
\label{fig:sequence_top_edge}
\end{figure*}

\begin{figure*}[ht!]
\centering
\includegraphics[width=\linewidth]{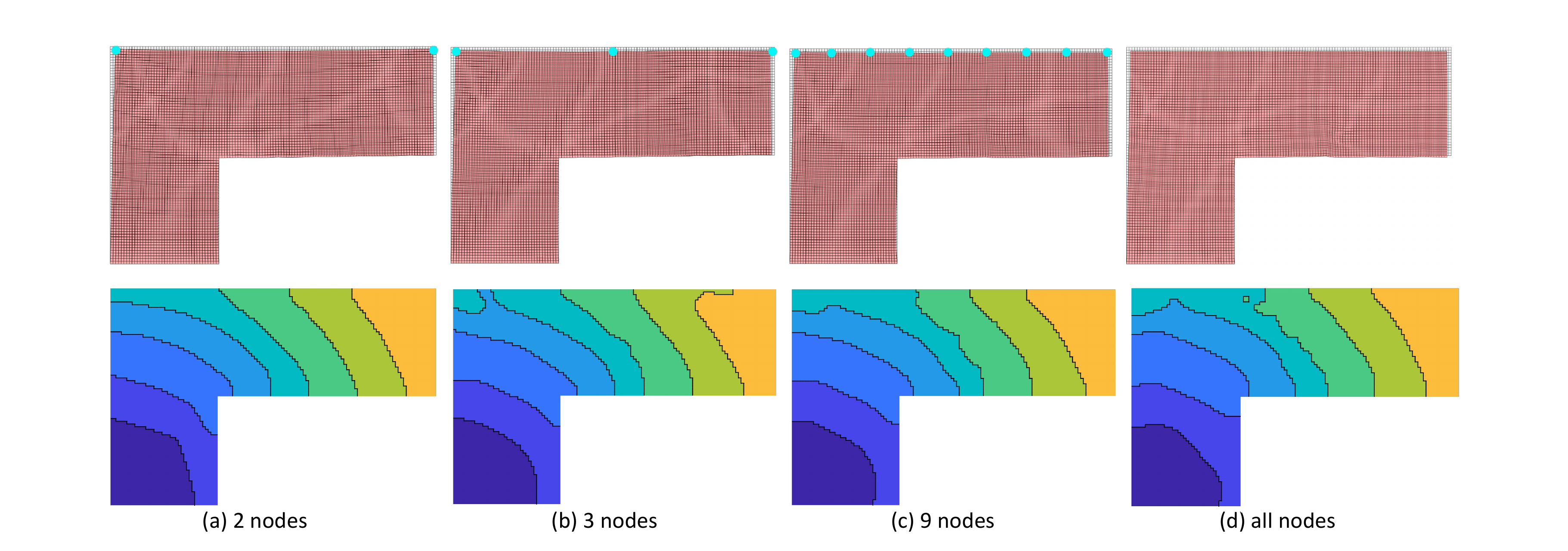}
\caption{Optimized sequences (bottom) and the corresponding simulated distortion (top) for maintaining the flatness of the top edge, with the flatness being defined by the vertical mismatch of, from left to right, 2, 3, 9, and all nodes on the top edge.}
\label{fig:top_edge_more_nodes}
\end{figure*}

Figure~\ref{fig:sequence_top_edge} shows process simulation using the optimized curved layers, corresponding to Test-2 in Fig.~\ref{fig:Tests}. The entire domain of the component, including both solid elements (i.e., already fabricated part) and the empty grid, is simulated in each step. The empty grid passively deforms along with the already fabricated part. While during the first few steps, the top edge in the passive part deviates from a horizontal line, it gradually aligns horizontally in the last steps.

Figure~\ref{fig:top_edge_more_nodes} shows distortion with the flatness of the top edge defined by different numbers of sample nodes. The sample nodes are equally spaced on the top edge. Compared to Test-2 in Fig.~\ref{fig:Tests}c, as more nodes are introduced to define flatness, the bend in the middle of the edge is gradually flattened.

\begin{figure*}[ht!]
\centering
\includegraphics[width=\linewidth]{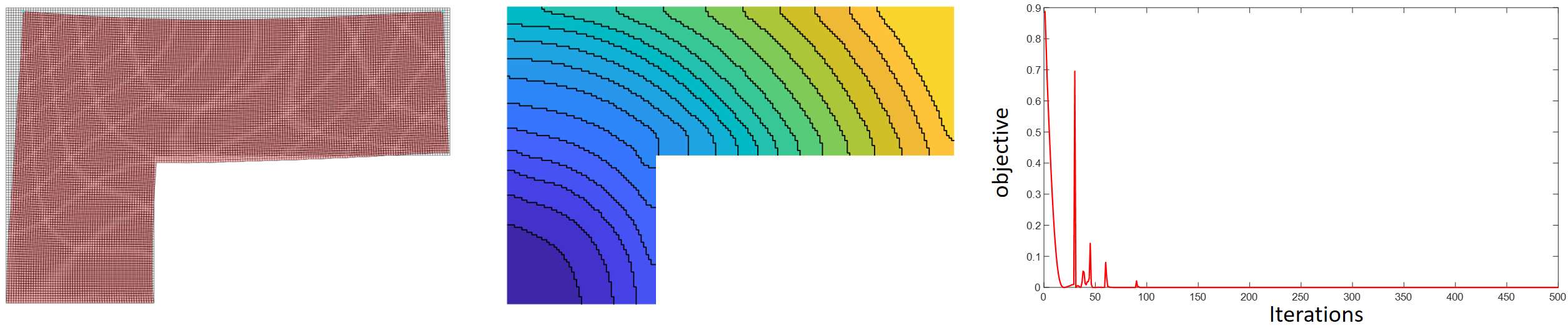}
\caption{The simulated distortion (left) and optimized fabrication sequence (middle), tested on a higher resolution and with 20 layers. The history of the objective value over iterations is plotted on the right. }
\label{fig:high_resolution}
\end{figure*}

We further validate our method with a higher resolution and a larger number of layers. This is shown in Fig.~\ref{fig:high_resolution} where the resolution is $180\times 120$, and the number of layers is set to $20$. The results are comparable to the one shown in Fig.~\ref{fig:Tests}c. The plot shown in Fig.~\ref{fig:high_resolution} (right) confirms the convergence of the gradient-based numerical optimization. There are some oscillations on the convergence curve due to a continuation of $\beta_t$ in Eq.~(\ref{eq:projection}). It starts from 30 and is increased by 10 every 30 iterations, until it reaches 100. The same setting is used in all 2D tests.

\subsubsection{Parameter study}

\begin{figure*}[ht!]
\centering
\includegraphics[width=\linewidth]{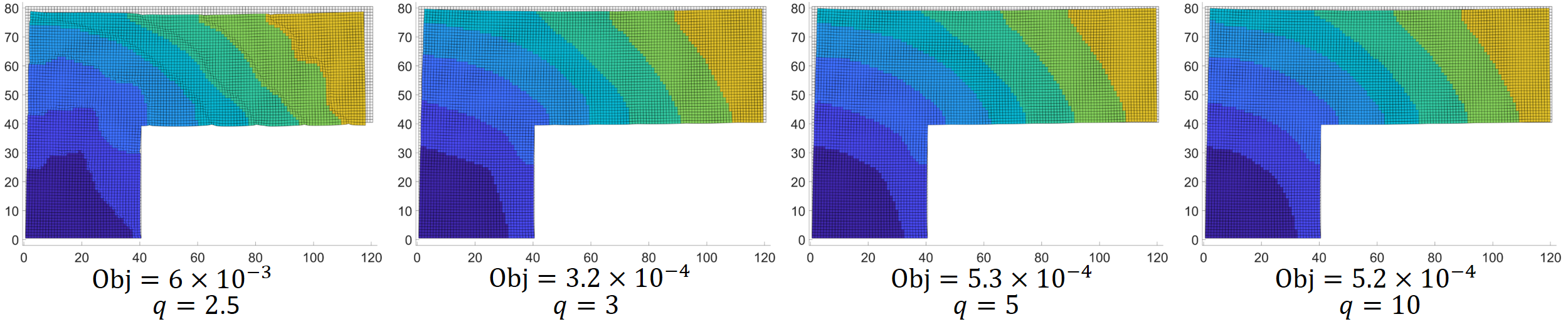}
\caption{Distortion minimization using different thermal-strain penalties ($q$ in Eq.~(\ref{eq:loads_proj})), and the same stiffness penalty ($p=3$). From left to right: $q < p$, $q = p$, and $q > p$.}
\label{fig:different_k}
\end{figure*}

In Fig.~\ref{fig:different_k} we study the influences of the thermal-strain penalty, $q$ in Eq.~(\ref{eq:loads_proj}). The stiffness penalty, $p$ in Eq.~(\ref{eq:Young_m}), is set to $3$. When $q<p$, the internal force due to the thermal strain is disproportionally larger than the stiffness, causing large distortion in finite element simulation. This distortion can be observed in Fig.~\ref{fig:different_k} (left). It also leads to convergence problems. While the objective was reduced to a small value, one of the volume constraints in this test was not satisfied. In contrast, when $q\geq p$, stable optimization is observed. As $q$ changes from 5 to 10, the optimized results show only subtle differences.

\begin{figure}[ht!]
\centering
\includegraphics[width=0.8\linewidth]{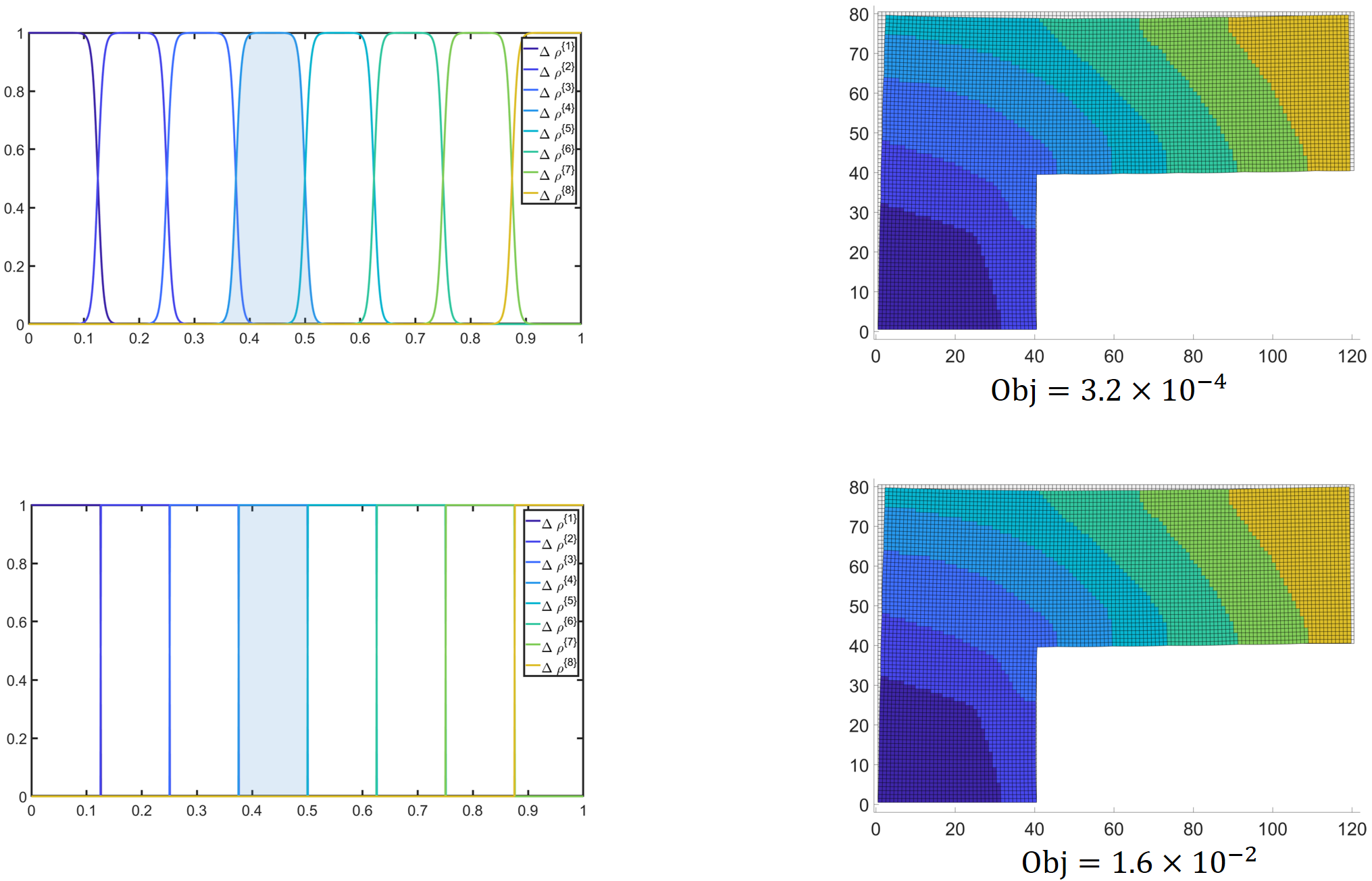}
\caption{Left: Illustration of the smoothed Heaviside projection (top) used in optimization and a binary 0-1 projection (bottom) for defining layers. Right: The simulated distortion using differentiable layers (top) and binary layers (bottom).}
\label{fig:simulate_realistic}
\end{figure}

Differential layers (i.e., with densities very close, but not equal to 0 or 1) have been used for simulating distortion in the optimization process. Fig.~\ref{fig:simulate_realistic} illustrates the smoothed Heaviside projection (top left) and a strict 0-1 projection (bottom left). The function corresponding to the 4-th layer is shaded for comparison.
To investigate the error due to this approximation, we perform distortion simulation with both differential layers and binary layers. The simulation results shown on the right-hand side suggest that the overall distortion is visually comparable. From the objective value, it is found that the measured distortion is larger when using the binary layers in a post-validation. This discrepancy could be reduced by increasing the sharpness of the smoothed Heaviside projection. Reflecting on the distortion measure (Eq.~(\ref{eq:obj-Test2})), it can be calculated that the vertical mismatch is $0.18$, i.e., less than a fifth of the unit length of a finite element or $0.15\%$ of the length of the component. We consider this a rather small distortion.

\subsection{2D bracket}

\begin{figure*}[ht!]
\centering
\includegraphics[width=\linewidth]{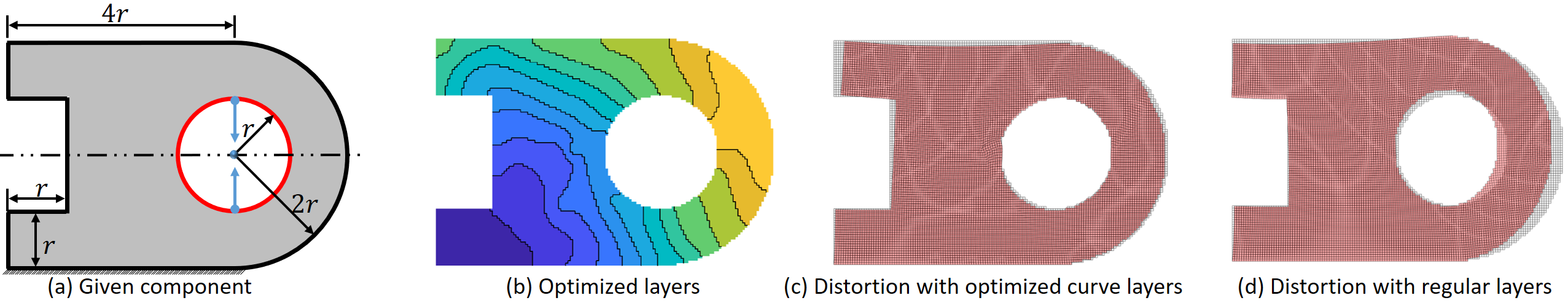}
\caption{The fabrication sequence of a 2D bracket (a) is optimized to minimize the average displacement of nodes on the inner circle (red). The fabrication starts from the bottom-left corner. The objective is reduced to $0.768$ (c) from $10.814$ (d).}
\label{fig:result_complex_structure}
\end{figure*}

The second example is a 2D bracket, shown in Fig.~\ref{fig:result_complex_structure}a. The domain is discretized by a finite element grid of $144 \times 96$. The fabrication is supposed to start from the corner on the bottom left of the component. The nodes on the bottom of the component are fixed in the process simulation. The number of layers is set to 12. The distortion is measured by the average of the displacements of nodes on the inner circle (red),
\begin{equation}
    d = \frac{1}{{n}(\mathcal{W})}{\sum_{j \in \mathcal{W}}} \left( \left(u^x_j\right)^2 + \left( u^y_j \right)^2 \right),
\end{equation}
where $\mathcal{W}$ denotes the set of nodes along the circle.

Figure~\ref{fig:result_complex_structure}b shows the optimized fabrication sequence. The resulting distortion is depicted in (c), while the distortion from regular horizontal layers is shown in (d) for comparison. The distortion in (d) makes the inner circle elliptic, while in (c) a circular hole is maintained at its intended location.

\begin{figure*}[ht!]
\centering
\includegraphics[width=0.9\linewidth]{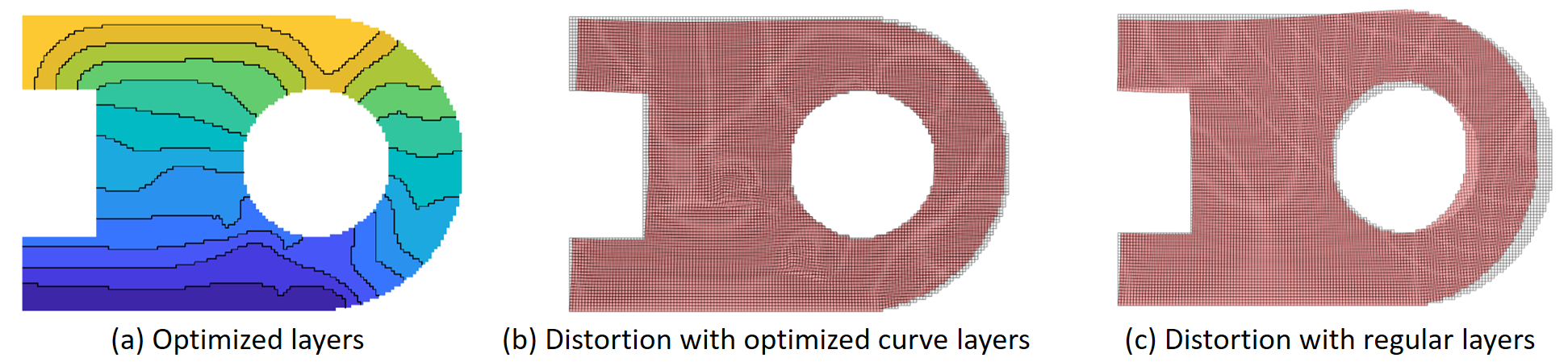}
\caption{(a) The optimized fabrication sequence where the fabrication is supposed to start from the bottom edge of the component. The objective is reduced to $0.207$ (b) from $10.814$ (c).}
\label{fig:result_complex_structure2}
\end{figure*}

We further test the influence of the prescribed starting region of fabrication on the optimized distortion. In Fig.~\ref{fig:result_complex_structure2}a, the fabrication starts from the bottom edge of the 2D shape. Fig.~\ref{fig:result_complex_structure2}b depicts the optimized distortion. The predicted distortion is 0.207. Comparing Fig.~\ref{fig:result_complex_structure2} with Fig.~\ref{fig:result_complex_structure}, it can be found that while the algorithm works effectively with different starting regions, the starting region has a large influence on the final distortion. Finding an optimal starting region thus can be an interesting topic for further research.

\subsection{3D components}

\begin{figure*}[ht!]
\centering
\includegraphics[width=\linewidth]{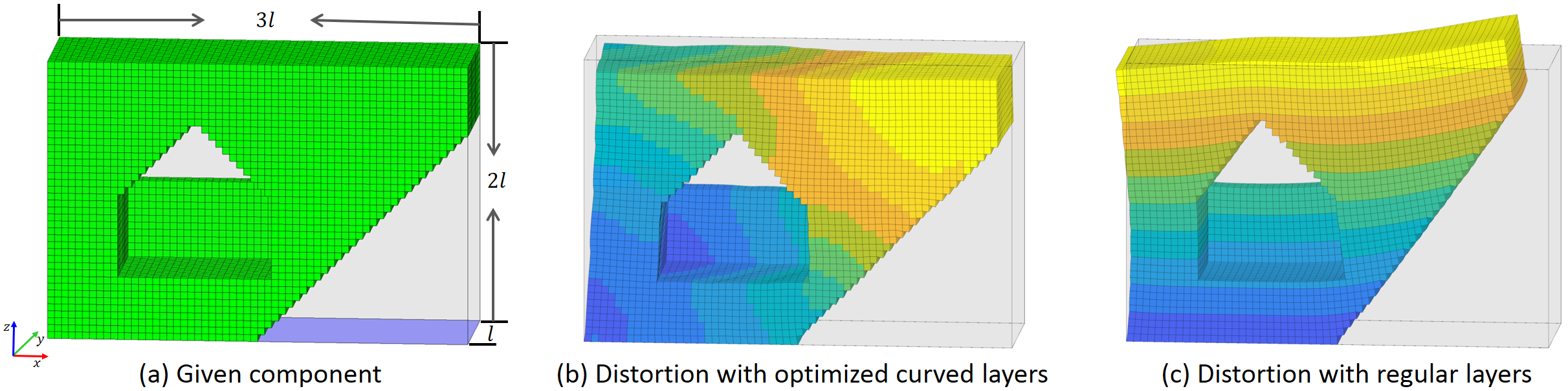}
\caption{Sequence optimization is performed to maintain the flatness and orientation of the surfaces on the right and on the top of the given component~(a). In process simulation, nodes on the bottom surface (blue) are assumed to be fixed. The fabrication is assumed to start from a corner on the bottom. The optimized layers and final distortion can be seen in (b) while the distortion of planar layers is shown in (c).}
\label{fig:LShape_3D}
\end{figure*}

\begin{figure*}[ht!]
\centering
\includegraphics[width=\linewidth]{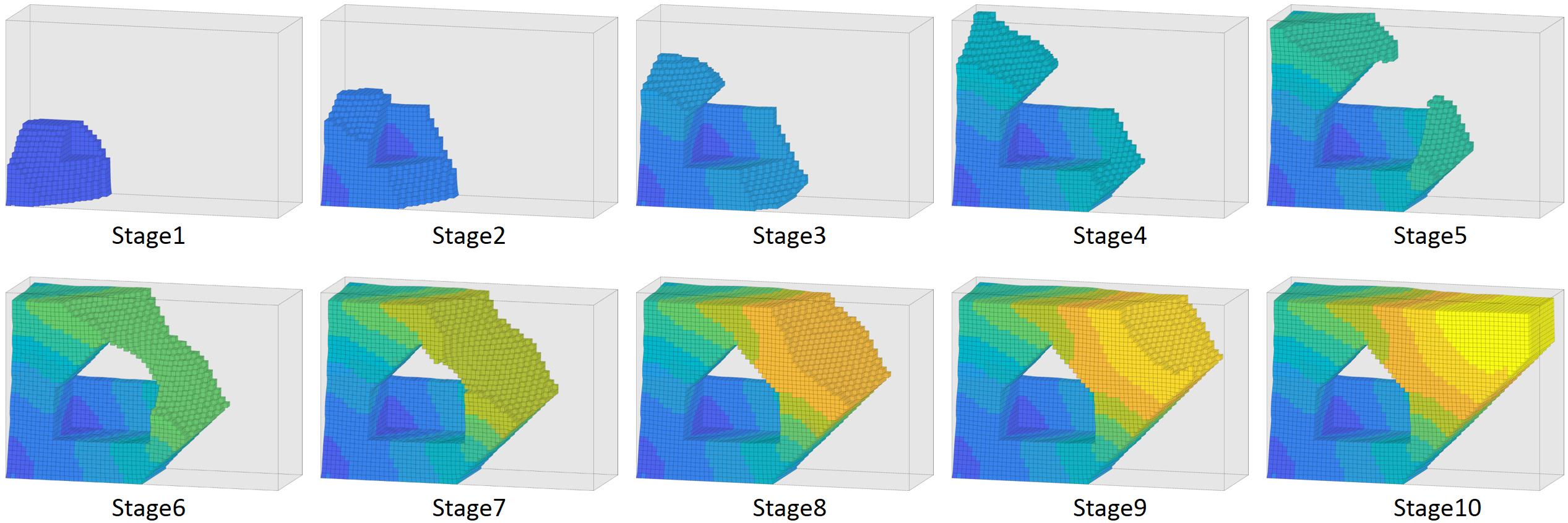}
\caption{The mechanical simulation of the fabrication process using optimized curved layers shown in Fig.~\ref{fig:LShape_3D}b.}
\label{fig:LShape_3D_sequence}
\end{figure*}

Our framework is directly applicable to 3D scenarios. Fig.~\ref{fig:LShape_3D}a illustrates the domain of a 3D component. It is discretized by a finite element grid of $60 \times 20 \times 40$. This component is optimized with 10 layers. The distortion is measured by the flatness and orientation of the surfaces on the right and on the top, by calculating the mismatch of nodes along the $x$-axis and $z$-axis, respectively,
\begin{equation}
    d = \frac{1}{{n}(\mathcal{I}_r)}\sum_{i \in \mathcal{I}_r}\left(u^x_i - \textstyle{\frac{1}{{n}(\mathcal{I}_r)}} \textstyle{\sum_{i \in \mathcal{I}_r}} u^x_i \right)^2 + \frac{1}{{n}(\mathcal{I}_t)}\sum_{i \in \mathcal{I}_t}\left(u^z_i - \textstyle{\frac{1}{{n}(\mathcal{I}_t)}} \textstyle{\sum_{i \in \mathcal{I}_t}} u^z_i \right)^2,
    \label{eq:obj_3D}
\end{equation}
where $\mathcal{I}_r$ and $\mathcal{I}_t$ are the sets of nodes sampled on the right and top surfaces, respectively.
A comparison of distortion from optimized curved layers and regular planar layers can be seen in Fig.~\ref{fig:LShape_3D} (b) and (c). The quantitative measure of distortion is $0.1173$ (b) vs. $1.7075$ (c). Fig.~\ref{fig:LShape_3D_sequence} illustrates the simulated sequence of fabrication using the optimized layers.

\begin{figure*}[ht!]
\centering
\includegraphics[width=\linewidth]{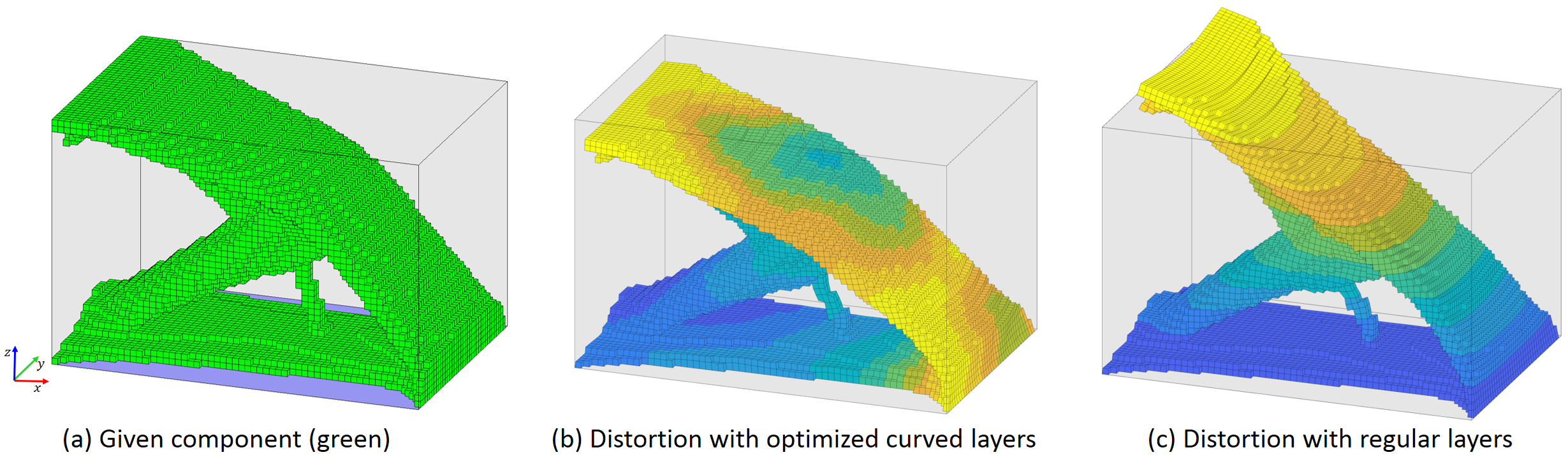}
\caption{The given component (a) is designed by topology optimization. Its fabrication sequence is then optimized to maintain the flatness and orientation of the surfaces on the left and on the top. The distortion is reduced to $0.0267$ (b) from $0.9888$ (c).}
\label{fig:Topopt_3D}
\end{figure*}

\begin{figure*}[ht!]
\centering
\includegraphics[width=\linewidth]{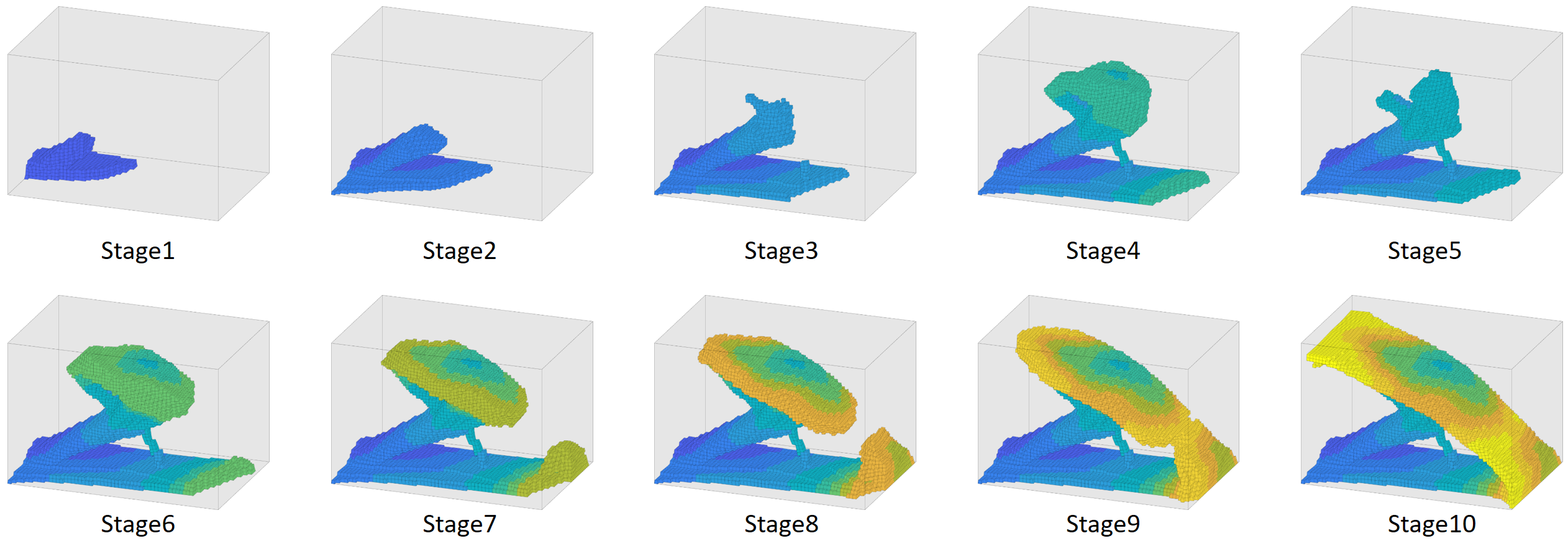}
\caption{The mechanical simulation of the fabrication process using optimized curved layers shown in Fig.~\ref{fig:Topopt_3D}b.}
\label{fig:Topopt_3D_sequence}
\end{figure*}

The last example is a 3D structure designed by topology optimization (Fig.~\ref{fig:Topopt_3D}a). The domain of the structure is discretized by a finite element grid of $60 \times 40 \times 40$. The optimized structure has a volume fraction of $0.2$. It serves as input to our sequence optimization. For process simulation, the bottom of the domain is fixed, and one of its corners is prescribed as the startpoint of the fabrication. The fabrication sequence for this component is optimized with 10 layers. The objective function is the flatness of the surfaces on the top and on the left (i.e., similar to Eq.~(\ref{eq:obj_3D})). The simulated distortion with optimized layers is shown in Fig.~\ref{fig:Topopt_3D}b. As a comparison, the distortion from regular horizontal layers is shown in Fig.~\ref{fig:Topopt_3D}c. The quantitative measure of distortion is $0.0267$ (b) vs. $0.9888$ (c). The simulation sequence with optimized curved layers is visualized in Fig.~\ref{fig:Topopt_3D_sequence}.

\subsection{Anisotropic inherent strain}
\label{subsec:anisotropy}

To demonstrate the extendability of our framework, we replace the isotropic inherent strain with an anisotropic one. In 2D, the anisotropic strain $\underline{\bm{\varepsilon}}^* = [\epsilon_x, \epsilon_y, \epsilon_{xy}]^{\T}$ is set as $\epsilon_x=-0.01$, $\epsilon_y=0$, and $\epsilon_{xy}=0$. Here, the subscripts $x$ and $y$ refer to a local coordinate system associated with each element. The $x$-axis represents the material deposition direction at the element. The orthogonal $y$-axis is defined by the gradient of the time field. To facilitate the computation of the gradient, we define the time field on the nodes of the finite element grid. The gradient at each element is then calculated by the bilinear shape function. We use the normalized gradient of the time field to transform the anisotropic inherent strain to the global coordinate system for finite element analysis. This transformation is incorporated into the calculation of the nodal forces that are equivalent to the inherent strain (Eq.~\ref{eq:nodalForces}). Accordingly, the sensitivity analysis needs to consider the dependence of the nodal forces on the time field, Eq.~\ref{eq:forceGradient} in particular.

\begin{figure*}[ht!]
\centering
\includegraphics[width=\linewidth]{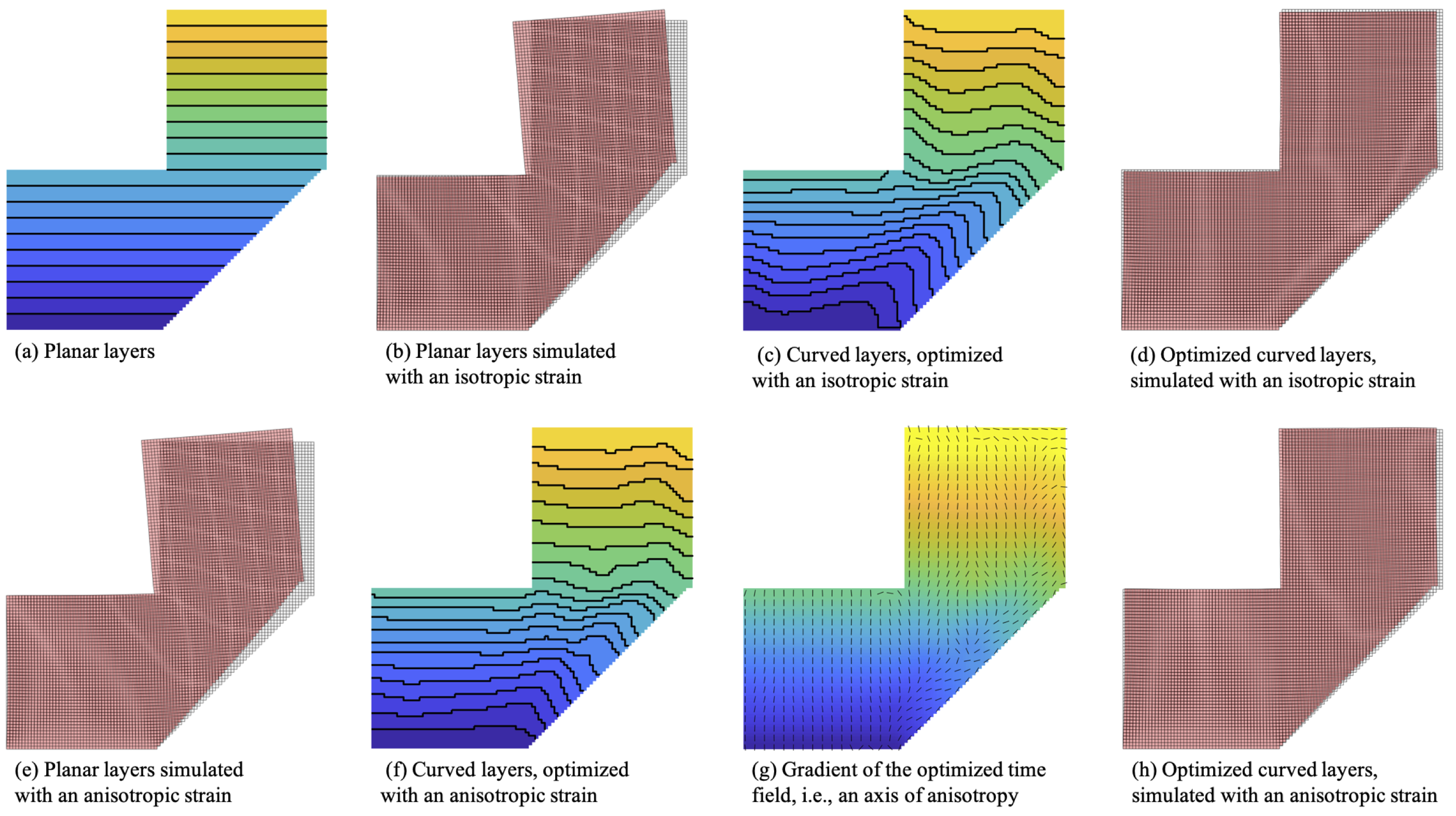}
\caption{Fabrication sequence optimization of a 2D component using an isotropic inherent strain (top row) and an anisotropic one (bottom row). The objective is to keep the top edge of the component horizontal. In both cases, the distortion is significantly reduced. In the isotropic case, the distortion is reduced from $2.57$ of the planar layers~(b) to $8.85\times10^{-3}$ of the optimized curved layers~(d). In the anisotropic case, an axis of the anisotropic strain is aligned with the gradient of the time field~(g). The distortion is reduced from $2.57$ of the planar layers~(e) to $2.42\times10^{-3}$ of the optimized curved layers~(h).}
\label{fig:anistropicInherentStrain}
\end{figure*}

Figure~\ref{fig:anistropicInherentStrain} shows the results of fabrication sequence optimization of a 2D component using an isotropic inherent strain (top row) and an anisotropic one (bottom row). The 2D domain is discretized using a finite element grid of $100\times100$. The number of layers is $20$. The objective is to keep the top edge of the component horizontal (Eq.~\ref{eq:obj-Test2}), measured on three vertices, i.e., at each end of the edge, and at its middle. When planar layers~(a) are used, the flatness of the top edge is the same for the isotropic and anisotropic inherent strain (b and e, respectively). However, the average of the vertical displacements is different, $1.49$~(b) and $2.49$~(e). The average of vertical displacements is reduced to $-0.80$~(d) and $0.41$~(h), respectively. In both cases, the distortion objective is significantly reduced, from $2.57$ to $8.85\times10^{-3}$~(d) and $2.42\times10^{-3}$~(h). This validates the effectiveness of the proposed fabrication sequence optimization framework for different process models.

\subsection{Computational complexity}

The computational complexity of the proposed framework is linear with respect to the (prescribed) number of layers. Compared to the classic compliance minimization problem which requires solving a single state equation, here we need to solve $N$ state equations, with $N$ being the number of layers. Each state equation corresponds to the deposition of a new layer (cf. Eq.~\ref{eq:equilibriumLayer}). Additionally, for each state equation, the sensitivity analysis in distortion minimization involves solving the adjoint state equation (cf. Eq.~\ref{eq:adjoint}). Thus, roughly speaking, the complexity is about $2N$ times that of the commonly solved compliance minimization problem. Under the assumption of small deformation, the state equations are independent~\cite{munro2019process} and can thus be solved in parallel, though in our implementation we did not explore this. As an indication of the computational time, sequence optimization of the 2D $L$-shaped component with a finite element grid of $120\times80$ and 8 layers in Figs.~\ref{fig:Tests},~\ref{fig:top_edge_more_nodes} and~\ref{fig:different_k} took 15.5 minutes on average. The higher resolution 2D $L$-shaped component with a finite element grid of $180\times120$ and 20 layers took 84.8 minutes. The running time of the 3D examples shown in Figs.~\ref{fig:LShape_3D} and~\ref{fig:Topopt_3D} was slightly less than 15 hours. The tests were based on an implementation using Matlab, on a PC equipped with Intel(R) Core(TM) i7-6700K CPU @ 4.00GHz and 64GB memory. A single CPU core was used. All results shown in this work were generated with 500 optimization iterations.

The number of layers is a prescribed parameter in our framework. It can be selected by
\begin{equation}
    N = \frac{V^{*}}{\bar{h}\bar{l}},
\end{equation}
where $V^{*}$ is the total volume of the component. $\bar{h}$ is the desired nominal thickness of layers. $\bar{l}$ is the average length of layers, which can be estimated based on the dimension of the component.
\\

\section{Conclusions and Future Work}
\label{sec:conclusion}

In this paper, we have presented a computational framework for fabrication sequence optimization to reduce distortion in multi-axis additive manufacturing (e.g., robotic wire arc additive manufacturing). The optimization takes a pseudo-time field as design variables to mimic the fabrication sequence. The time field segments the given component, in a differentiable manner, into (curved) layers. Numerical results show that the fabrication sequence has a huge influence on the distortion. By optimizing the sequence, the distortion is reduced by more than an order of magnitude in comparison to commonly-used planar layers. The improved dimensional accuracy is especially relevant for large metal components. The optimization framework is not restricted to a certain distortion model, while in this paper it is demonstrated using inherent strains (both isotropic and anisotropic) to simulate distortion.

As future work, we intend to carry out an experimental investigation, first to calibrate the thermal strain, and then to validate the effectiveness of the optimized sequence on distortion minimization. The reduction in distortion is likely accompanied by an increase in residual stresses. In parallel to physical tests, we plan to integrate a restriction on thermal stresses during the additive process into our sequence optimization framework. Stress is a local property, and the stress field evolves during fabrication. This means a very large number of constraints has to be integrated into the optimization. 

The proposed computational framework opens up a new dimension in exploring the full potential of additive manufacturing. The current consideration of manufacturability is by no means comprehensive. Firstly, as can be observed in e.g. Fig.~\ref{fig:LShape_3D}, the layers have varying thicknesses. By varying the material deposition rate and movement speed of the print head~\cite{Kuipers2020CAD}, the layer thickness can be adjusted during fabrication, but only to a limited extent. An idea to reduce the variation is by imposing minimum and maximum length scales on each individual layer. Here the robust formulation~\cite{Wang2010SMO} may serve as a good starting point. Secondly, while the rotation of the build plate offers a possibility to greatly reduce or even eliminate support structures~\cite{dai2018support}, it does not guarantee support-free fabrication for complex parts with complex curved layers. The geometric condition for support-free fabrication needs to be further investigated, and to be incorporated into fabrication sequence optimization. Thirdly, the rotation of the build plate potentially gives rise to interference between the print head and the already deposited part. This issue is coupled with support-free process planning. A potential solution to this is to restrict the orientation of curved layers in the optimization.

\small{\textbf{Acknowledgements} The authors gratefully acknowledge the support from the LEaDing Fellows Programme at the Delft University of Technology, which has received funding from the European Union's Horizon 2020 research and innovation programme under the Marie Skłodowska-Curie grant agreement No. 707404. Weiming Wang wishes to thank the National Natural Science Foundation of China (No. 62172073), and the Natural Science Foundation of Liaoning Province (No. 2021-MS-110).}

\bibliography{mybib}

\begin{thebibliography}{44}
\expandafter\ifx\csname natexlab\endcsname\relax\def\natexlab#1{#1}\fi
\providecommand{\url}[1]{\texttt{#1}}
\providecommand{\href}[2]{#2}
\providecommand{\path}[1]{#1}
\providecommand{\DOIprefix}{doi:}
\providecommand{\ArXivprefix}{arXiv:}
\providecommand{\URLprefix}{URL: }
\providecommand{\Pubmedprefix}{pmid:}
\providecommand{\doi}[1]{\href{http://dx.doi.org/#1}{\path{#1}}}
\providecommand{\Pubmed}[1]{\href{pmid:#1}{\path{#1}}}
\providecommand{\bibinfo}[2]{#2}
\ifx\xfnm\relax \def\xfnm[#1]{\unskip,\space#1}\fi
\bibitem[{Williams et~al.(2016)Williams, Martina, Addison, Ding, Pardal, and
  Colegrove}]{williams2016wire+}
\bibinfo{author}{S.~W. Williams}, \bibinfo{author}{F.~Martina},
  \bibinfo{author}{A.~C. Addison}, \bibinfo{author}{J.~Ding},
  \bibinfo{author}{G.~Pardal}, \bibinfo{author}{P.~Colegrove},
\newblock \bibinfo{title}{Wire+ arc additive manufacturing},
\newblock \bibinfo{journal}{Materials Science and Technology}
  \bibinfo{volume}{32} (\bibinfo{year}{2016}) \bibinfo{pages}{641--647}.
  \DOIprefix\doi{10.1179/1743284715Y.0000000073}.
\bibitem[{Ding et~al.(2011)Ding, Colegrove, Mehnen, Ganguly, Almeida, Wang, and
  Williams}]{ding2011thermo}
\bibinfo{author}{J.~Ding}, \bibinfo{author}{P.~Colegrove},
  \bibinfo{author}{J.~Mehnen}, \bibinfo{author}{S.~Ganguly},
  \bibinfo{author}{P.~S. Almeida}, \bibinfo{author}{F.~Wang},
  \bibinfo{author}{S.~Williams},
\newblock \bibinfo{title}{Thermo-mechanical analysis of wire and arc additive
  layer manufacturing process on large multi-layer parts},
\newblock \bibinfo{journal}{Computational Materials Science}
  \bibinfo{volume}{50} (\bibinfo{year}{2011}) \bibinfo{pages}{3315--3322}.
  \DOIprefix\doi{10.1016/j.commatsci.2011.06.023}.
\bibitem[{Dai et~al.(2018)Dai, Wang, Wu, Lefebvre, Fang, and
  Liu}]{dai2018support}
\bibinfo{author}{C.~Dai}, \bibinfo{author}{C.~C. Wang},
  \bibinfo{author}{C.~Wu}, \bibinfo{author}{S.~Lefebvre},
  \bibinfo{author}{G.~Fang}, \bibinfo{author}{Y.-J. Liu},
\newblock \bibinfo{title}{Support-free volume printing by multi-axis motion},
\newblock \bibinfo{journal}{ACM Transactions on Graphics} \bibinfo{volume}{37}
  (\bibinfo{year}{2018}) \bibinfo{pages}{1--14}.
  \DOIprefix\doi{10.1145/3197517.3201342}.
\bibitem[{Fang et~al.(2020)Fang, Zhang, Zhong, Chen, Zhong, and
  Wang}]{Fang2020ToG}
\bibinfo{author}{G.~Fang}, \bibinfo{author}{T.~Zhang},
  \bibinfo{author}{S.~Zhong}, \bibinfo{author}{X.~Chen},
  \bibinfo{author}{Z.~Zhong}, \bibinfo{author}{C.~C.~L. Wang},
\newblock \bibinfo{title}{Reinforced fdm: Multi-axis filament alignment with
  controlled anisotropic strength},
\newblock \bibinfo{journal}{ACM Trans. Graph.} \bibinfo{volume}{39}
  (\bibinfo{year}{2020}). \DOIprefix\doi{10.1145/3414685.3417834}.
\bibitem[{Wang et~al.(2020)Wang, Munro, Wang, van Keulen, and
  Wu}]{wang2020space}
\bibinfo{author}{W.~Wang}, \bibinfo{author}{D.~Munro}, \bibinfo{author}{C.~C.
  Wang}, \bibinfo{author}{F.~van Keulen}, \bibinfo{author}{J.~Wu},
\newblock \bibinfo{title}{Space-time topology optimization for additive
  manufacturing},
\newblock \bibinfo{journal}{Structural and Multidisciplinary Optimization}
  \bibinfo{volume}{61} (\bibinfo{year}{2020}) \bibinfo{pages}{1--18}.
  \DOIprefix\doi{10.1007/s00158-019-02420-6}.
\bibitem[{Jafari et~al.(2021)Jafari, Vaneker, and Gibson}]{jafari2021wire}
\bibinfo{author}{D.~Jafari}, \bibinfo{author}{T.~H. Vaneker},
  \bibinfo{author}{I.~Gibson},
\newblock \bibinfo{title}{Wire and arc additive manufacturing: Opportunities
  and challenges to control the quality and accuracy of manufactured parts},
\newblock \bibinfo{journal}{Materials \& Design}  (\bibinfo{year}{2021})
  \bibinfo{pages}{109471}. \DOIprefix\doi{10.1016/j.matdes.2021.109471}.
\bibitem[{Zalameda et~al.(2013)Zalameda, Burke, Hafley, Taminger, Domack,
  Brewer, and Martin}]{zalameda2013thermal}
\bibinfo{author}{J.~N. Zalameda}, \bibinfo{author}{E.~R. Burke},
  \bibinfo{author}{R.~A. Hafley}, \bibinfo{author}{K.~M. Taminger},
  \bibinfo{author}{C.~S. Domack}, \bibinfo{author}{A.~Brewer},
  \bibinfo{author}{R.~E. Martin},
\newblock \bibinfo{title}{Thermal imaging for assessment of electron-beam
  freeform fabrication (ebf3) additive manufacturing deposits},
\newblock in: \bibinfo{booktitle}{Thermosense: Thermal Infrared Applications
  XXXV}, volume \bibinfo{volume}{8705}, \bibinfo{organization}{International
  Society for Optics and Photonics}, \bibinfo{year}{2013}, p.
  \bibinfo{pages}{87050M}. \DOIprefix\doi{10.1117/12.2018233}.
\bibitem[{Mueller and Mueller(2000)}]{mueller2000experiences}
\bibinfo{author}{D.~Mueller}, \bibinfo{author}{H.~Mueller},
\newblock \bibinfo{title}{Experiences using rapid prototyping techniques to
  manufacture sheet metal forming tools},
\newblock in: \bibinfo{booktitle}{Proc. ISATA Conference, Dublin},
  volume~\bibinfo{volume}{9}, \bibinfo{year}{2000}.
  \DOIprefix\doi{10.1.1.122.2370}.
\bibitem[{Ya and Hamilton(2017)}]{ya2017demand}
\bibinfo{author}{W.~Ya}, \bibinfo{author}{K.~Hamilton},
\newblock \bibinfo{title}{On-demand spare parts for the marine industry with
  directed energy deposition: propeller use case},
\newblock in: \bibinfo{booktitle}{International Conference on Additive
  Manufacturing in Products and Applications},
  \bibinfo{organization}{Springer}, \bibinfo{year}{2017}, pp.
  \bibinfo{pages}{70--81}. \DOIprefix\doi{10.1007/978-3-319-66866-6_7}.
\bibitem[{Nycz et~al.(2017)Nycz, Noakes, Richardson, Messing, Post, Paul,
  Flamm, and Love}]{nycz2017challenges}
\bibinfo{author}{A.~Nycz}, \bibinfo{author}{M.~W. Noakes},
  \bibinfo{author}{B.~Richardson}, \bibinfo{author}{A.~Messing},
  \bibinfo{author}{B.~Post}, \bibinfo{author}{J.~Paul},
  \bibinfo{author}{J.~Flamm}, \bibinfo{author}{L.~Love},
\newblock \bibinfo{title}{Challenges in making complex metal large-scale parts
  for additive manufacturing: A case study based on the additive manufacturing
  excavator},
\newblock in: \bibinfo{booktitle}{2017 International Solid Freeform Fabrication
  Symposium}, \bibinfo{organization}{University of Texas at Austin},
  \bibinfo{year}{2017}. \DOIprefix\doi{10.26153/16923}.
\bibitem[{Nycz et~al.(2021)Nycz, Lee, Noakes, Ankit, Masuo, Simunovic, Bunn,
  Love, Oancea, Payzant, and Fancher}]{nycz2021effective}
\bibinfo{author}{A.~Nycz}, \bibinfo{author}{Y.~Lee},
  \bibinfo{author}{M.~Noakes}, \bibinfo{author}{D.~Ankit},
  \bibinfo{author}{C.~Masuo}, \bibinfo{author}{S.~Simunovic},
  \bibinfo{author}{J.~Bunn}, \bibinfo{author}{L.~Love},
  \bibinfo{author}{V.~Oancea}, \bibinfo{author}{A.~Payzant},
  \bibinfo{author}{C.~M. Fancher},
\newblock \bibinfo{title}{Effective residual stress prediction validated with
  neutron diffraction method for metal large-scale additive manufacturing},
\newblock \bibinfo{journal}{Materials \& Design} \bibinfo{volume}{205}
  (\bibinfo{year}{2021}) \bibinfo{pages}{109751}.
  \DOIprefix\doi{10.1016/j.matdes.2021.109751}.
\bibitem[{Romero-Hdz et~al.(2016)Romero-Hdz, Saha, Toledo-Ramirez, and
  Beltran-Bqz}]{romero2016welding}
\bibinfo{author}{J.~Romero-Hdz}, \bibinfo{author}{B.~Saha},
  \bibinfo{author}{G.~Toledo-Ramirez}, \bibinfo{author}{D.~Beltran-Bqz},
\newblock \bibinfo{title}{Welding sequence optimization using artificial
  intelligence techniques: an overview},
\newblock \bibinfo{journal}{Int J Comput Sci Eng} \bibinfo{volume}{3}
  (\bibinfo{year}{2016}) \bibinfo{pages}{90--95}.
  \DOIprefix\doi{10.14445/23488387/IJCSE-V3I11P115}.
\bibitem[{Beik et~al.(2019)Beik, Marzbani, and Jazar}]{beik2019welding}
\bibinfo{author}{V.~Beik}, \bibinfo{author}{H.~Marzbani},
  \bibinfo{author}{R.~Jazar},
\newblock \bibinfo{title}{Welding sequence optimisation in the automotive
  industry: A review},
\newblock \bibinfo{journal}{Proceedings of the Institution of Mechanical
  Engineers, Part C: Journal of Mechanical Engineering Science}
  \bibinfo{volume}{233} (\bibinfo{year}{2019}) \bibinfo{pages}{5945--5952}.
  \DOIprefix\doi{10.1177/0954406219859909}.
\bibitem[{Sattari-Far and Javadi(2008)}]{sattari2008influence}
\bibinfo{author}{I.~Sattari-Far}, \bibinfo{author}{Y.~Javadi},
\newblock \bibinfo{title}{Influence of welding sequence on welding distortions
  in pipes},
\newblock \bibinfo{journal}{International Journal of Pressure Vessels and
  Piping} \bibinfo{volume}{85} (\bibinfo{year}{2008})
  \bibinfo{pages}{265--274}. \DOIprefix\doi{10.1016/j.ijpvp.2007.07.003}.
\bibitem[{Ding et~al.(2015)Ding, Pan, Cuiuri, and Li}]{ding2015wire}
\bibinfo{author}{D.~Ding}, \bibinfo{author}{Z.~Pan},
  \bibinfo{author}{D.~Cuiuri}, \bibinfo{author}{H.~Li},
\newblock \bibinfo{title}{Wire-feed additive manufacturing of metal components:
  technologies, developments and future interests},
\newblock \bibinfo{journal}{The International Journal of Advanced Manufacturing
  Technology} \bibinfo{volume}{81} (\bibinfo{year}{2015})
  \bibinfo{pages}{465--481}. \DOIprefix\doi{10.1007/s00170-015-7077-3}.
\bibitem[{Mughal et~al.(2005)Mughal, Fawad, Mufti, and
  Siddique}]{mughal2005deformation}
\bibinfo{author}{M.~Mughal}, \bibinfo{author}{H.~Fawad},
  \bibinfo{author}{R.~Mufti}, \bibinfo{author}{M.~Siddique},
\newblock \bibinfo{title}{Deformation modelling in layered manufacturing of
  metallic parts using gas metal arc welding: effect of process parameters},
\newblock \bibinfo{journal}{Modelling and Simulation in Materials Science and
  Engineering} \bibinfo{volume}{13} (\bibinfo{year}{2005})
  \bibinfo{pages}{1187}. \DOIprefix\doi{10.1088/0965-0393/13/7/013}.
\bibitem[{Mughal et~al.(2007)Mughal, Mufti, and Fawad}]{mughal2007mechanical}
\bibinfo{author}{M.~Mughal}, \bibinfo{author}{R.~Mufti},
  \bibinfo{author}{H.~Fawad},
\newblock \bibinfo{title}{The mechanical effects of deposition patterns in
  welding-based layered manufacturing},
\newblock \bibinfo{journal}{Proceedings of the Institution of Mechanical
  Engineers, Part B: Journal of Engineering Manufacture} \bibinfo{volume}{221}
  (\bibinfo{year}{2007}) \bibinfo{pages}{1499--1509}.
  \DOIprefix\doi{10.1243/09544054JEM783}.
\bibitem[{Graf et~al.(2018)Graf, H{\"a}lsig, H{\"o}fer, Awiszus, and
  Mayr}]{graf2018thermo}
\bibinfo{author}{M.~Graf}, \bibinfo{author}{A.~H{\"a}lsig},
  \bibinfo{author}{K.~H{\"o}fer}, \bibinfo{author}{B.~Awiszus},
  \bibinfo{author}{P.~Mayr},
\newblock \bibinfo{title}{Thermo-mechanical modelling of wire-arc additive
  manufacturing ({WAAM}) of semi-finished products},
\newblock \bibinfo{journal}{Metals} \bibinfo{volume}{8} (\bibinfo{year}{2018})
  \bibinfo{pages}{1009}. \DOIprefix\doi{10.3390/met8121009}.
\bibitem[{Yang et~al.(2016)Yang, Zhang, Cheng, Min, Chyu, and
  To}]{yang2016finite}
\bibinfo{author}{Q.~Yang}, \bibinfo{author}{P.~Zhang},
  \bibinfo{author}{L.~Cheng}, \bibinfo{author}{Z.~Min},
  \bibinfo{author}{M.~Chyu}, \bibinfo{author}{A.~C. To},
\newblock \bibinfo{title}{Finite element modeling and validation of
  thermomechanical behavior of ti-6al-4v in directed energy deposition additive
  manufacturing},
\newblock \bibinfo{journal}{Additive Manufacturing} \bibinfo{volume}{12}
  (\bibinfo{year}{2016}) \bibinfo{pages}{169--177}.
  \DOIprefix\doi{10.1016/j.addma.2016.06.012}.
\bibitem[{K{\"o}rner et~al.(2020)K{\"o}rner, Markl, and Koepf}]{Korner2020MMTA}
\bibinfo{author}{C.~K{\"o}rner}, \bibinfo{author}{M.~Markl},
  \bibinfo{author}{J.~A. Koepf},
\newblock \bibinfo{title}{Modeling and simulation of microstructure evolution
  for additive manufacturing of metals: a critical review},
\newblock \bibinfo{journal}{Metallurgical and Materials Transactions A}
  \bibinfo{volume}{51} (\bibinfo{year}{2020}) \bibinfo{pages}{4970--4983}.
  \DOIprefix\doi{10.1007/s11661-020-05946-3}.
\bibitem[{Wei et~al.(2021)Wei, Mukherjee, Zhang, Zuback, Knapp, De, and
  DebRoy}]{Wei2021PMS}
\bibinfo{author}{H.~Wei}, \bibinfo{author}{T.~Mukherjee},
  \bibinfo{author}{W.~Zhang}, \bibinfo{author}{J.~Zuback},
  \bibinfo{author}{G.~Knapp}, \bibinfo{author}{A.~De},
  \bibinfo{author}{T.~DebRoy},
\newblock \bibinfo{title}{Mechanistic models for additive manufacturing of
  metallic components},
\newblock \bibinfo{journal}{Progress in Materials Science}
  \bibinfo{volume}{116} (\bibinfo{year}{2021}) \bibinfo{pages}{100703}.
  \DOIprefix\doi{10.1016/j.pmatsci.2020.100703}.
\bibitem[{Singh et~al.(2022)Singh, M.Singari, and Mishra}]{Singh2022MTP}
\bibinfo{author}{P.~Singh}, \bibinfo{author}{R.~M.Singari},
  \bibinfo{author}{R.~Mishra},
\newblock \bibinfo{title}{A review of study on modeling and simulation of
  additive manufacturing processes},
\newblock \bibinfo{journal}{Materials Today: Proceedings} \bibinfo{volume}{56}
  (\bibinfo{year}{2022}) \bibinfo{pages}{3594--3603}.
  \DOIprefix\doi{10.1016/j.matpr.2021.12.057}, \bibinfo{note}{first
  International Conference on Design and Materials}.
\bibitem[{Denlinger and Michaleris(2017)}]{denlinger2017mitigation}
\bibinfo{author}{E.~R. Denlinger}, \bibinfo{author}{P.~Michaleris},
\newblock \bibinfo{title}{Mitigation of distortion in large additive
  manufacturing parts},
\newblock \bibinfo{journal}{Proceedings of the Institution of Mechanical
  Engineers, Part B: Journal of Engineering Manufacture} \bibinfo{volume}{231}
  (\bibinfo{year}{2017}) \bibinfo{pages}{983--993}.
  \DOIprefix\doi{10.1177/0954405415578580}.
\bibitem[{Denlinger and Michaleris(2016)}]{denlinger2016effect}
\bibinfo{author}{E.~R. Denlinger}, \bibinfo{author}{P.~Michaleris},
\newblock \bibinfo{title}{Effect of stress relaxation on distortion in additive
  manufacturing process modeling},
\newblock \bibinfo{journal}{Additive Manufacturing} \bibinfo{volume}{12}
  (\bibinfo{year}{2016}) \bibinfo{pages}{51--59}.
  \DOIprefix\doi{10.1016/j.addma.2016.06.011}.
\bibitem[{Salonitis et~al.(2016)Salonitis, D’Alvise, Schoinochoritis, and
  Chantzis}]{salonitis2016additive}
\bibinfo{author}{K.~Salonitis}, \bibinfo{author}{L.~D’Alvise},
  \bibinfo{author}{B.~Schoinochoritis}, \bibinfo{author}{D.~Chantzis},
\newblock \bibinfo{title}{Additive manufacturing and post-processing
  simulation: laser cladding followed by high speed machining},
\newblock \bibinfo{journal}{The International Journal of Advanced Manufacturing
  Technology} \bibinfo{volume}{85} (\bibinfo{year}{2016})
  \bibinfo{pages}{2401--2411}. \DOIprefix\doi{10.1007/s00170-015-7989-y}.
\bibitem[{Chiumenti et~al.(2010)Chiumenti, Cervera, Salmi, De~Saracibar,
  Dialami, and Matsui}]{chiumenti2010finite}
\bibinfo{author}{M.~Chiumenti}, \bibinfo{author}{M.~Cervera},
  \bibinfo{author}{A.~Salmi}, \bibinfo{author}{C.~A. De~Saracibar},
  \bibinfo{author}{N.~Dialami}, \bibinfo{author}{K.~Matsui},
\newblock \bibinfo{title}{Finite element modeling of multi-pass welding and
  shaped metal deposition processes},
\newblock \bibinfo{journal}{Computer methods in applied mechanics and
  engineering} \bibinfo{volume}{199} (\bibinfo{year}{2010})
  \bibinfo{pages}{2343--2359}. \DOIprefix\doi{10.1016/j.cma.2010.02.018}.
\bibitem[{Ueda et~al.(1975)Ueda, Fukuda, Nakacho, and Endo}]{ueda1975new}
\bibinfo{author}{Y.~Ueda}, \bibinfo{author}{K.~Fukuda},
  \bibinfo{author}{K.~Nakacho}, \bibinfo{author}{S.~Endo},
\newblock \bibinfo{title}{A new measuring method of residual stresses with the
  aid of finite element method and reliability of estimated values},
\newblock \bibinfo{journal}{Journal of the Society of Naval Architects of
  Japan} \bibinfo{volume}{1975} (\bibinfo{year}{1975})
  \bibinfo{pages}{499--507}. \DOIprefix\doi{10.2534/jjasnaoe1968.1975.138_499}.
\bibitem[{Murakawa et~al.(1996)Murakawa, Luo, and
  Ueda}]{murakawa1996prediction}
\bibinfo{author}{H.~Murakawa}, \bibinfo{author}{Y.~Luo},
  \bibinfo{author}{Y.~Ueda},
\newblock \bibinfo{title}{Prediction of welding deformation and residual stress
  by elastic fem based on inherent strain},
\newblock \bibinfo{journal}{Journal of the society of Naval Architects of
  Japan} \bibinfo{volume}{1996} (\bibinfo{year}{1996})
  \bibinfo{pages}{739--751}. \DOIprefix\doi{10.2534/jjasnaoe1968.1996.180_739}.
\bibitem[{Ma et~al.(2016)Ma, Nakacho, Ohta, Ogawa, Maekawa, Huang, and
  Murakawa}]{ma2016inherent}
\bibinfo{author}{N.~Ma}, \bibinfo{author}{K.~Nakacho},
  \bibinfo{author}{T.~Ohta}, \bibinfo{author}{N.~Ogawa},
  \bibinfo{author}{A.~Maekawa}, \bibinfo{author}{H.~Huang},
  \bibinfo{author}{H.~Murakawa},
\newblock \bibinfo{title}{Inherent strain method for residual stress
  measurement and welding distortion prediction},
\newblock in: \bibinfo{booktitle}{International Conference on Offshore
  Mechanics and Arctic Engineering}, volume \bibinfo{volume}{50008},
  \bibinfo{organization}{American Society of Mechanical Engineers},
  \bibinfo{year}{2016}, p. \bibinfo{pages}{V009T13A001}.
  \DOIprefix\doi{10.1115/OMAE2016-54184}.
\bibitem[{Munro et~al.(2019)Munro, Ayas, Langelaar, and van
  Keulen}]{munro2019process}
\bibinfo{author}{D.~Munro}, \bibinfo{author}{C.~Ayas},
  \bibinfo{author}{M.~Langelaar}, \bibinfo{author}{F.~van Keulen},
\newblock \bibinfo{title}{On process-step parallel computability and linear
  superposition of mechanical responses in additive manufacturing process
  simulation},
\newblock \bibinfo{journal}{Additive Manufacturing} \bibinfo{volume}{28}
  (\bibinfo{year}{2019}) \bibinfo{pages}{738--749}.
  \DOIprefix\doi{10.1016/j.addma.2019.06.023}.
\bibitem[{Liang et~al.(2019)Liang, Chen, Cheng, Hayduke, and
  To}]{liang2019modified}
\bibinfo{author}{X.~Liang}, \bibinfo{author}{Q.~Chen},
  \bibinfo{author}{L.~Cheng}, \bibinfo{author}{D.~Hayduke},
  \bibinfo{author}{A.~C. To},
\newblock \bibinfo{title}{Modified inherent strain method for efficient
  prediction of residual deformation in direct metal laser sintered
  components},
\newblock \bibinfo{journal}{Computational Mechanics} \bibinfo{volume}{64}
  (\bibinfo{year}{2019}) \bibinfo{pages}{1719--1733}.
  \DOIprefix\doi{10.1007/s00466-019-01748-6}.
\bibitem[{Chen et~al.(2019)Chen, Liang, Hayduke, Liu, Cheng, Oskin, Whitmore,
  and To}]{chen2019inherent}
\bibinfo{author}{Q.~Chen}, \bibinfo{author}{X.~Liang},
  \bibinfo{author}{D.~Hayduke}, \bibinfo{author}{J.~Liu},
  \bibinfo{author}{L.~Cheng}, \bibinfo{author}{J.~Oskin},
  \bibinfo{author}{R.~Whitmore}, \bibinfo{author}{A.~C. To},
\newblock \bibinfo{title}{An inherent strain based multiscale modeling
  framework for simulating part-scale residual deformation for direct metal
  laser sintering},
\newblock \bibinfo{journal}{Additive Manufacturing} \bibinfo{volume}{28}
  (\bibinfo{year}{2019}) \bibinfo{pages}{406--418}.
  \DOIprefix\doi{10.1016/j.addma.2019.05.021}.
\bibitem[{Prabhune and Suresh(2020)}]{prabhune2020fast}
\bibinfo{author}{B.~C. Prabhune}, \bibinfo{author}{K.~Suresh},
\newblock \bibinfo{title}{A fast matrix-free elasto-plastic solver for
  predicting residual stresses in additive manufacturing},
\newblock \bibinfo{journal}{Computer-Aided Design} \bibinfo{volume}{123}
  (\bibinfo{year}{2020}) \bibinfo{pages}{102829}.
  \DOIprefix\doi{10.1016/j.cad.2020.102829}.
\bibitem[{Allaire and Jakab{\v{c}}in(2018)}]{allaire2018taking}
\bibinfo{author}{G.~Allaire}, \bibinfo{author}{L.~Jakab{\v{c}}in},
\newblock \bibinfo{title}{Taking into account thermal residual stresses in
  topology optimization of structures built by additive manufacturing},
\newblock \bibinfo{journal}{Mathematical Models and Methods in Applied
  Sciences} \bibinfo{volume}{28} (\bibinfo{year}{2018})
  \bibinfo{pages}{2313--2366}. \DOIprefix\doi{10.1142/S0218202518500501}.
\bibitem[{Pellens et~al.(2020)Pellens, Lombaert, Michiels, Craeghs, and
  Schevenels}]{pellens2020topology}
\bibinfo{author}{J.~Pellens}, \bibinfo{author}{G.~Lombaert},
  \bibinfo{author}{M.~Michiels}, \bibinfo{author}{T.~Craeghs},
  \bibinfo{author}{M.~Schevenels},
\newblock \bibinfo{title}{Topology optimization of support structure layout in
  metal-based additive manufacturing accounting for thermal deformations},
\newblock \bibinfo{journal}{Structural and Multidisciplinary Optimization}
  \bibinfo{volume}{61} (\bibinfo{year}{2020}) \bibinfo{pages}{2291--2303}.
  \DOIprefix\doi{10.1007/s00158-020-02512-8}.
\bibitem[{Misiun et~al.(2021)Misiun, van~de Ven, Langelaar, Geijselaers, van
  Keulen, van~den Boogaard, and Ayas}]{misiun2021topology}
\bibinfo{author}{G.~Misiun}, \bibinfo{author}{E.~van~de Ven},
  \bibinfo{author}{M.~Langelaar}, \bibinfo{author}{H.~Geijselaers},
  \bibinfo{author}{F.~van Keulen}, \bibinfo{author}{T.~van~den Boogaard},
  \bibinfo{author}{C.~Ayas},
\newblock \bibinfo{title}{Topology optimization for additive manufacturing with
  distortion constraints},
\newblock \bibinfo{journal}{Computer Methods in Applied Mechanics and
  Engineering} \bibinfo{volume}{386} (\bibinfo{year}{2021})
  \bibinfo{pages}{114095}. \DOIprefix\doi{10.1016/j.cma.2021.114095}.
\bibitem[{Miki and Yamada(2021)}]{Miki2021CMAME}
\bibinfo{author}{T.~Miki}, \bibinfo{author}{T.~Yamada},
\newblock \bibinfo{title}{Topology optimization considering the distortion in
  additive manufacturing},
\newblock \bibinfo{journal}{Finite Elements in Analysis and Design}
  \bibinfo{volume}{193} (\bibinfo{year}{2021}) \bibinfo{pages}{103558}.
  \DOIprefix\doi{10.1016/j.finel.2021.103558}.
\bibitem[{Wang et~al.(2010)Wang, Lazarov, and Sigmund}]{Wang2010SMO}
\bibinfo{author}{F.~Wang}, \bibinfo{author}{B.~S. Lazarov},
  \bibinfo{author}{O.~Sigmund},
\newblock \bibinfo{title}{On projection methods, convergence and robust
  formulations in topology optimization},
\newblock \bibinfo{journal}{Struct. Multidiscip. Optim.} \bibinfo{volume}{43}
  (\bibinfo{year}{2010}) \bibinfo{pages}{767--784}.
  \DOIprefix\doi{10.1007/s00158-010-0602-y}.
\bibitem[{Wu et~al.(2018)Wu, Aage, Westermann, and Sigmund}]{Wu2018TVCG}
\bibinfo{author}{J.~Wu}, \bibinfo{author}{N.~Aage},
  \bibinfo{author}{R.~Westermann}, \bibinfo{author}{O.~Sigmund},
\newblock \bibinfo{title}{Infill optimization for additive manufacturing --
  approaching bone-like porous structures},
\newblock \bibinfo{journal}{IEEE Transactions on Visualization and Computer
  Graphics} \bibinfo{volume}{24} (\bibinfo{year}{2018})
  \bibinfo{pages}{1127--1140}. \DOIprefix\doi{10.1109/TVCG.2017.2655523}.
\bibitem[{Setien et~al.(2019)Setien, Chiumenti, van~der Veen, San~Sebastian,
  Garciand{\'\i}a, and Echeverr{\'\i}a}]{setien2019empirical}
\bibinfo{author}{I.~Setien}, \bibinfo{author}{M.~Chiumenti},
  \bibinfo{author}{S.~van~der Veen}, \bibinfo{author}{M.~San~Sebastian},
  \bibinfo{author}{F.~Garciand{\'\i}a}, \bibinfo{author}{A.~Echeverr{\'\i}a},
\newblock \bibinfo{title}{Empirical methodology to determine inherent strains
  in additive manufacturing},
\newblock \bibinfo{journal}{Computers \& Mathematics with Applications}
  \bibinfo{volume}{78} (\bibinfo{year}{2019}) \bibinfo{pages}{2282--2295}.
  \DOIprefix\doi{10.1016/j.camwa.2018.05.015}.
\bibitem[{Liang et~al.(2018)Liang, Cheng, Chen, Yang, and To}]{Liang2018AM}
\bibinfo{author}{X.~Liang}, \bibinfo{author}{L.~Cheng},
  \bibinfo{author}{Q.~Chen}, \bibinfo{author}{Q.~Yang}, \bibinfo{author}{A.~C.
  To},
\newblock \bibinfo{title}{A modified method for estimating inherent strains
  from detailed process simulation for fast residual distortion prediction of
  single-walled structures fabricated by directed energy deposition},
\newblock \bibinfo{journal}{Additive Manufacturing} \bibinfo{volume}{23}
  (\bibinfo{year}{2018}) \bibinfo{pages}{471--486}.
  \DOIprefix\doi{10.1016/j.addma.2018.08.029}.
\bibitem[{Michaleris(2014)}]{michaleris2014modeling}
\bibinfo{author}{P.~Michaleris},
\newblock \bibinfo{title}{Modeling metal deposition in heat transfer analyses
  of additive manufacturing processes},
\newblock \bibinfo{journal}{Finite Elements in Analysis and Design}
  \bibinfo{volume}{86} (\bibinfo{year}{2014}) \bibinfo{pages}{51--60}.
  \DOIprefix\doi{10.1016/j.finel.2014.04.003}.
\bibitem[{Svanberg(1987)}]{svanberg1987method}
\bibinfo{author}{K.~Svanberg},
\newblock \bibinfo{title}{The method of moving asymptotes—a new method for
  structural optimization},
\newblock \bibinfo{journal}{International journal for numerical methods in
  engineering} \bibinfo{volume}{24} (\bibinfo{year}{1987})
  \bibinfo{pages}{359--373}. \DOIprefix\doi{10.1002/nme.1620240207}.
\bibitem[{Kuipers et~al.(2020)Kuipers, Doubrovski, Wu, and
  Wang}]{Kuipers2020CAD}
\bibinfo{author}{T.~Kuipers}, \bibinfo{author}{E.~L. Doubrovski},
  \bibinfo{author}{J.~Wu}, \bibinfo{author}{C.~C. Wang},
\newblock \bibinfo{title}{A framework for adaptive width control of dense
  contour-parallel toolpaths in fused deposition modeling},
\newblock \bibinfo{journal}{Computer-Aided Design} \bibinfo{volume}{128}
  (\bibinfo{year}{2020}) \bibinfo{pages}{102907}.
  \DOIprefix\doi{10.1016/j.cad.2020.102907}.

\end{thebibliography}

\end{document}